\documentclass[journal]{IEEEtran}
\IEEEoverridecommandlockouts
\usepackage{amsmath}
\usepackage{amsfonts}
\usepackage{amssymb}
\usepackage[english]{babel}
\usepackage{amsthm}

\usepackage{esint}
\usepackage{algorithm,algorithmicx}
\usepackage{algpseudocode}
\usepackage{array}
\usepackage[caption=false,font=normalsize,labelfont=sf,textfont=sf]{subfig}
\usepackage{textcomp}
\usepackage{stfloats}
\usepackage{multirow}
\usepackage{booktabs}
\usepackage{etoolbox}
\usepackage{geometry}
\usepackage{pstricks}

\usepackage{svg}

\usepackage{url}
\usepackage{verbatim}

\usepackage{cite}
\def\BibTeX{{\rm B\kern-.05em{\sc i\kern-.025em b}\kern-.08em
    T\kern-.1667em\lower.7ex\hbox{E}\kern-.125emX}}
\usepackage{subfloat}
\usepackage{tikz}
\usetikzlibrary{graphs, arrows, positioning, quotes, shapes.geometric, arrows.meta}

\ifCLASSINFOpdf
\else
\fi

\begin{document}

\title{\Huge Metasurface-Enabled Superheterodyne Transmitter With Decoupled Harmonic-Free Signal Generation and Precoding}

\author{
Xuehui~Dong,~\IEEEmembership{Student Member,~IEEE,}
Miyu~Feng,
Chen~Shao,
Bokai~Lai,
Jianan~Zhang,~\IEEEmembership{Student Member,~IEEE,}
Rujing~Xiong,~\IEEEmembership{Member,~IEEE,}
Kai~Wan,~\IEEEmembership{Member,~IEEE,} 
Tiebin~Mi,~\IEEEmembership{Member,~IEEE,} 
and~Robert~Caiming~Qiu,~\IEEEmembership{Fellow,~IEEE}

\thanks{X.~Dong, M.~Feng, C.~Shao, B.~Lai, J.~Zhang, K.~Wan, T.~Mi and R.~C.~Qiu are  with the School of Electronic Information and Communications, 
Huazhong University of Science and Technology, 430074  Wuhan, China,  (e-mail: \{xuehuidong, miyu\_feng, shaochen0517, bokailai,  kai\_wan, zhangjn, mitiebin, caiming\}@hust.edu.cn).
R.~Xiong are with the School of Science and Engineering, The Chinese University of Hong Kong, Shenzhen, 518172, China.
(e-mail: rujingxiong@cuhk.edu.cn).
(\textit{Corresponding author: Robert Caiming Qiu})
}
}

\markboth{IEEE Journal of \LaTeX\ Class Files,~Vol.~14, No.~8, July~2025}%
{Shell \MakeLowercase{\textit{et al.}}: Bare Demo of IEEEtran.cls for IEEE Journals}

\maketitle

\begin{abstract}
The evolution of programmable metasurfaces (PM) from passive beamforming to active information transmission marks a paradigm shift for next-generation wireless systems. However, this transition is hindered by fundamental limitations in conventional metasurface transmitter architectures, including restricted modulation orders, symbol-level spatial inconsistency, and significant harmonic interference. These issues stem from the intrinsic coupling between baseband signal processing and radio-frequency beamforming in monolithic designs reliant on simplistic switching mechanisms. This paper proposes a novel metasurface-enabled superheterodyne architecture (MSA) that fundamentally decouples these functionalities. The MSA introduces a dual-stage up-conversion process, comprising a digital up-conversion module for in-phase/quadrature modulation and baseband-to-intermediate frequency conversion, a precoder module for precoding, and a custom-designed magnitude-phase-decoupled metasurface that acts as a reconfigurable reflective mixer array. This decoupling of harmonic-free waveform generation from spatial precoding overcomes the critical drawbacks of existing approaches. Experimental results from a 5.8 GHz proof-of-concept prototype system validate the MSA's superior performance. The system generates spatially isotropic constellations for arbitrary-order QAM modulations, ensures consistent time-frequency signatures for applications like Doppler-spoofing, and achieves data rates up to 20 Mbps within a linear operating region that minimizes nonlinear distortion. The capability of employing spatial diversity and multi-stream interference cancellation has been demonstrated for the first time in a PM-based transmitter.

\end{abstract}


\IEEEpeerreviewmaketitle

\section{Introduction}
In conventionally envisioned wireless environments, programmable metasurfaces (PM), also known as reconfigurable intelligent surfaces (RIS), are integrated into the environment to reconstruct wireless channels by manipulating the propagation of electromagnetic (EM) waves, thereby enhancing the performance of existing communication systems~\cite{renzo2019smart,wu2019towards,strinati2021reconfigurable}. However, with the development of metasurfaces and metamaterials~\cite{cui2024roadmap,ma2022information,maci2024electromagnetic}, simply deploying a large number of metasurfaces to reconstruct wireless channels can no longer meet the demands of future wireless communication systems. Therefore, how to endow metasurfaces with more functionalities has become one of the current research hotspots. 

The core merit of PM is their programmable, dynamic control over EM properties such as amplitude, phase, and polarization. Although early studies had a predominant focus on beamforming for improved coverage, their functionality has since evolved beyond wave manipulation to include sensing, wireless power transfer, and imaging, representing a trend towards multifunctional systems. Envision a sixth-generation (6G) communication network environment in which every surface—ranging from walls to vehicles—acts as an imperceptible yet efficient intelligent information source, facilitating seamless data transmission and active sensing~\cite{umer2025reconfigurable,tishchenko2025emergence}. This vision is brought within reach by multifunctional PM, which are envisioned to evolve from being `passive reflectors' that enhance link performance to becoming `active transmitters' capable of directly modulating and emitting information. However, this transition in their role exposes fundamental architectural limitations inherent to traditional metasurface-based transmitter designs based on simplistic switching mechanisms~\cite{han2024principles}.
\subsection{Existing Work}
Considerable research efforts have recently been devoted to exploring the potential of metasurfaces for information modulation and transmission.
Initially, metasurfaces were passive structures designed for wave manipulation based on the generalized Snell's law, enabling anomalous reflection and refraction~\cite{yu2011light}. These metasurfaces focused on phase, amplitude, and polarization control for applications like cloaking, holography, and lensing. For instance, static coding metasurfaces used geometric arrangements to achieve fixed beamforming patterns~\cite{cui2014coding}. However, their functionality was limited to analog wave control without programmability or information capacity. The introduction of digital coding in 2014 marked a turning point, where metasurface elements were represented by binary states (e.g., `0' and `1' corresponding to 0° and 180° phase shifts), enabling real-time reconfigurability through field-programmable gate arrays (FPGAs). The digital coding metasurfaces allowed dynamic control of EM waves through coding sequences, facilitating beam scanning and holographic imaging~\cite{liu2016convolution}. 
Building on digital coding, temporal coding emerged as the first scheme utilizing periodically varying digital control signals to encode information in the time or frequency domains, thereby generating specific time-frequency signatures~\cite{tang2019programmable,dai2019realization,dai2019wireless,cui2019direct}.
However, the modulation order in this scheme is fundamentally limited by diodes' number per unit cell, making high-order modulation challenging to implement in practice.

To overcome this limitation, space-time-coding was developed as an alternative approach~\cite{zhang2018space,zhang2021wireless}. This scheme employs joint optimization of coding sequences across spatial and temporal domains, enabling higher-order modulation that surpasses the fundamental constraints imposed by the limited number of diodes per unit cell. This advancement was further extended to harmonic control, where independent manipulation of harmonic amplitudes and phases allowed for advanced modulation schemes, such as 256 quadrature amplitude modulation (QAM), increasing data rates and spectral efficiency~\cite{chen2022accurate,dai2018independent}. Additionally, RISs were leveraged for backscatter communication, where metasurfaces reflected ambient signals while encoding data, reducing power consumption and enabling energy-efficient IoT devices~\cite{basar2019wireless,jiang2021long,liu2020vmscatter}. Beyond communication, information metasurfaces have been applied in radar deception and sensing scenarios, such as mimicking Doppler signatures for spoofing through custom time-frequency pattern synthesis~\cite{chen2014reduction,kozlov2023radar,wang2024radar}. Both these applications necessitate fundamental functionalities like precise signal generation and beamforming.

With their monolithic designs, metasurface-based transmitters are poised to underpin next-generation joint communication-sensing networks~\cite{zhang2021wireless}, embedding EM intelligence directly into radiating surfaces.
However, this architectural simplification introduces significant coupling between the baseband signal phase and the carrier phase. 
Consequently, the mapping from symbols to control codebooks becomes inconsistent across different metasurface configurations (e.g., size, quantization levels), incident wave conditions, and receiver locations.
This fundamental limitation originates from the inherent lack of hardware decoupling in existing metasurface-based transmitter architectures~\cite{tang2020wireless}.
The absence of conventional transmitter components (e.g., mixers, filters) necessitates implementing all transmitter functionalities through digital switching of nonlinear components (e.g., diodes, transistors) embedded in each unit cell.
Although recent studies have attempted to mitigate this entanglement and exploit it for user multiplexing by serving receivers at different locations, the harmonic components cannot be completely suppressed due to fundamental limitations imposed by Fourier analysis~\cite{luo2024fully,zhang2021wireless}.

Based on the architectural coupling issues outlined above, practical implementations of metasurface transmitters face further constraints at the component and system levels. Time-coding metasurfaces are limited by the finite number of diodes per unit cell, imposing a modulation-order ceiling and restricting achievable data rates. space-time-coding metasurfaces suffer from symbol anisotropy: the tight coupling of modulation and beamforming produces spatially varying constellations, so multipath components can carry inconsistent symbols and interfere at the receiver. Furthermore, both approaches rely on abrupt switching of nonlinear components, which generates significant harmonic and spurious emissions that degrade communication performance and undermine the fidelity of radar-deception waveforms. 
\subsection{Contributions}
This paper proposes a novel metasurface-enabled superheterodyne architecture (MSA) transmitter which enables harmonic-free modulation of arbitrary signals onto an incident RF carrier while independently controlling the beam pattern of the reflected wave. 
This decoupling allows the MSA to supports spatially isotropic transmission for arbitrary-order QAM modulations. This inherent programmability allows the system to exploit spatial diversity for link reliability and perform independent multi-stream spatial multiplexing for enhanced capacity. Furthermore, it empowers the generation of high-fidelity arbitrary time-frequency features, thereby seamlessly integrating sensing functionalities with communication.
The innovation of the MSA lies in its hardware-decoupled architecture, which constitutes a paradigm shift from conventional monolithic designs. It comprises three key modules:
\begin{itemize}
    \item The magnitude-phase-decoupled (MPD) metasurface features programmable unit cells that independently control magnitude and phase of the reflection coefficients (RCs). The time-varying magnitudes are responsible for carrying information, while the phases are employed for passive precoding.    
    \item The digital up-conversion (DUC) module handles all signal generation and baseband processing, including I/Q modulation, baseband-to-IF conversion, filtering, and digital-to-analog conversion.
    \item The precoder module computes the optimal codebooks for each unit cell based on the desired beamforming direction and the characteristics of the incident wave.
\end{itemize}
Operationally, the complex baseband signal is first up-converted to an intermediate frequency (IF) by the DUC module. This IF signal is then fed to the MPD-metasurface, which performs the final up-conversion to RF and simultaneous precoding. This dual-stage up-conversion process, inspired by superheterodyne principles, uniquely delegates the critical RF conversion and beamforming to the metasurface.

In summary, we make the following contributions:
\begin{itemize}
    \item We introduce a novel MSA that decouples baseband processing from RF beamforming. This represents a paradigm shift from monolithic designs and effectively overcomes limitations in modulation order, symbol anisotropy, and harmonic interference that plague conventional metasurface transmitters.
    \item We develop a comprehensive analytical framework comprising 1) a unit-level reflection model, 2) an array-level radiation model, and 3) an end-to-end communication model to theoretically demonstrate the principle of symbol-level spatial isotropy and to devise the corresponding precoding schemes for achieving spatial diversity and multi-stream interference cancellation.
    \item We design and implement a proof-of-concept prototype, featuring a 5.8 GHz MPD-metasurface, a DAC-based DUC module and a precoder module. The prototype demonstrates the feasibility and effectiveness of the proposed architecture in practical scenarios. The unit cell features a dual-via coupling structure and a dual-path configuration with integrated PIN diodes and two single-pole double-throw (SPDT) switches, enabling independent control of reflection magnitude and phase. 
    \item We provide extensive experimental validation in both communication and sensing scenarios. Experimental results demonstrate: 1) distortion-free mixing of arbitrary waveforms while maintaining stable radiation patterns; 2) spatially isotropic transmission of arbitrary-order QAM modulation; 3) effective harmonic interference mitigation through filtering; 4) robust spatial diversity and multiplexing in multipath environments; and 5) high-fidelity time-frequency signature generation. We additionally assess essential performance limits, such as the linear operational region for minimal distortion.
    \end{itemize}
\subsection{Organization of This Paper}
Section~\ref{Metasurface-enabled Superheterodyne Architecture} provides details of the MSA transmitter design, including the hardware prerequisite, decoupling principle and the superheterodyne design.
Section~\ref{Modeling of MSA} establishes a comprehensive analytical framework for the proposed MSA transmitter. Section~\ref{Functional Analysis} provides a thorough functional analysis and proposed the precoding schemes of spatial diversity and interference cancellation. Section~\ref{MPD Metasurface Design} presents the hardware design of MPD-metasurface unit. Section~\ref{Experimental Validation} validates the MSA's performance through extensive experiments. 
\section{Metasurface-enabled Superheterodyne Architecture\label{Metasurface-enabled Superheterodyne Architecture}}
\begin{figure*}[ht]
    \centering
        \includegraphics[width=\linewidth]{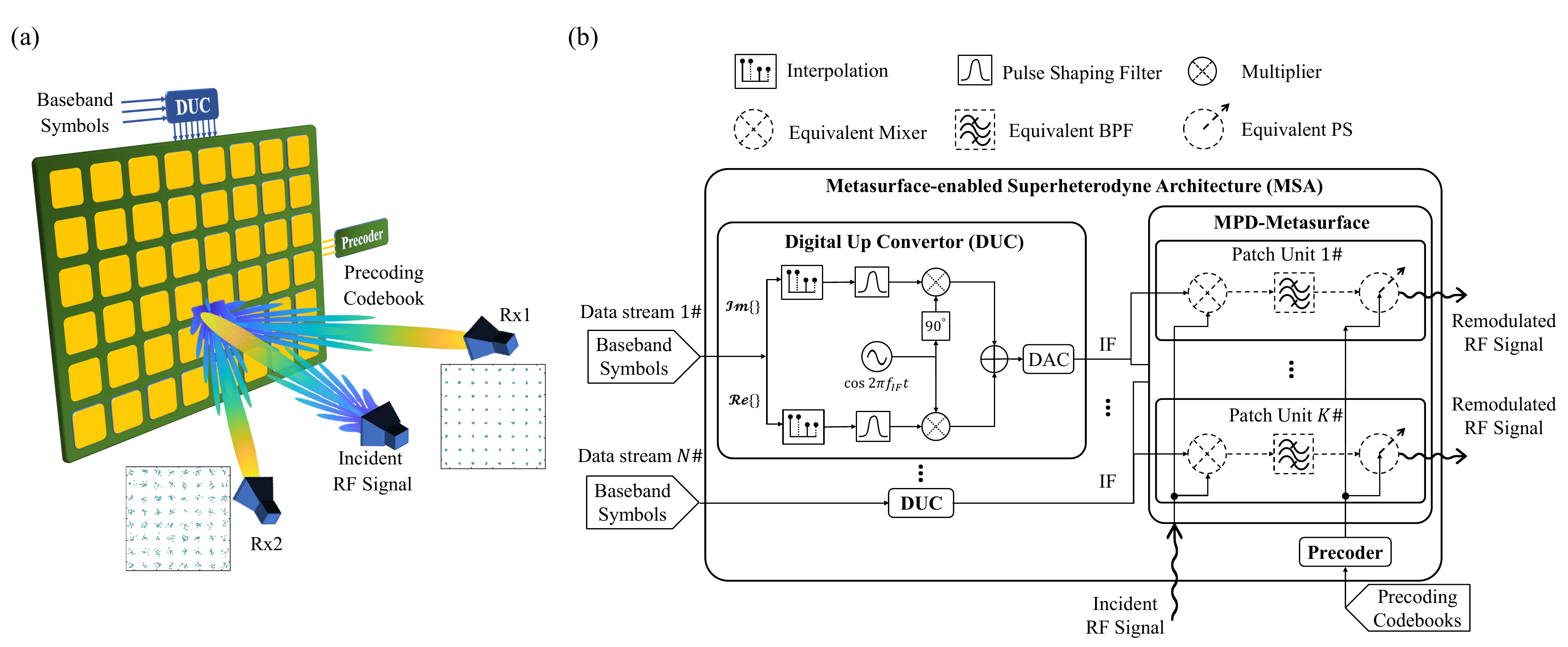}
        \caption{Conceptual illustration and Schematic diagram of MSA. (a) The conceptual illustration proposed MSA backscatter transmitter can independently generate and transmit arbitrary high-order complex signals while performing beamforming, ensuring spatial isotropy of the symbols. (b) The schematic diagram of the proposed MSA transmitter. }
    \label{Procedure}
\end{figure*}
Conventional metasurface transmitters adopt monolithic architectures that implement all functionalities through the switching of a limited set of discrete reflection states in the unit cells. 
While these designs are straightforward to realize, they inherently couple baseband signal processing with RF beamforming. 
This entanglement necessitates a joint codebook design for simultaneous information delivery and beamforming, which becomes heavily dependent on the specific propagation environment and physical configuration of the metasurface. 
Consequently, it leads to inherent symbol anisotropy and imposes a significant signal processing burden. 
Furthermore, the abrupt voltage transients from digitally switching the diodes introduce nonlinear distortion, generating unwanted harmonic interference that degrades the desired signal. 
These limitations collectively hinder the practical deployment and standardization of metasurface-based transmitters in real-world wireless systems.
To overcome these challenges, this section presents a novel MSA. 
\subsection{Hardware Prerequisite}
The core design philosophy of the MSA is functional decoupling, which separates baseband processing and beam steering into two distinct modules: the DUC module and the precoder module, as illustrated in Fig.~\ref{Procedure}(a). Similar to traditional multi-antenna transmitters, this architecture renders baseband signal generation independent of beamforming and radiation, thereby significantly simplifying signal processing and enhancing system flexibility.
As depicted in Fig.~\ref{Procedure}(b), the MSA comprises three key components:
\begin{itemize}
    \item \textbf{MPD-metasurface:} Serving as the innovative core of the MSA, this component operates as a reflective mixer and beamformer. It accepts an analog IF signal from the DUC module along with independent codebook from the precoder module. Its primary function is to mix the incident RF signal with the IF signal to conduct the final up-conversion, while simultaneously precoding the mixed signal toward the intended direction.
    \item \textbf{DUC module:} This module handles baseband-to-IF conversion, including I/Q modulation, filtering, and digital-to-analog conversion (DAC). It generates the analog IF complex signal that feeds the MPD-metasurface, enabling flexible modulation schemes and signal designs through separate baseband processing.
    \item \textbf{Precoder module}: This component generates steering commands for the MPD-metasurface based on the desired precoding codebook. Operating independently from baseband signal generation, it facilitates dynamic beam steering without affecting the modulation process.
\end{itemize}
From a system perspective, this implementation follows a superheterodyne architecture via a dual-stage up-conversion process: the DUC module first translates the complex baseband symbols to an IF signal through digital interpolation, pulse shaping, and I/Q modulation, and then up-converts it to the RF domain via the metasurface-based mixing operation.

Unlike traditional PM whose RC is controlled by a single variable (e.g., diode bias voltage), the MPD-metasurface features unit cells that independently control both the magnitude and phase of the RC through unique structural design. Each unit cell employs two control voltages: a continuous voltage for magnitude adjustment and a discrete voltage for phase tuning. Information is encoded in the amplitude of the reflected EM waves, while beam steering is achieved by programming the phase distribution across the metasurface via the precoder module.

As in Fig.~\ref{Procedure}(b), each unit simultaneously performs three integrated functions during RF conversion: (1) frequency mixing between incident RF signals and input IF signals, (2) out-of-band rejection through spectral filtering, and (3) programmable phase shifting of the outgoing mixed RF signal. The linearity of the magnitude-voltage response ensures minimal distortion in the mixing process, guaranteeing high-purity upconverted RF signals capable of supporting high-order modulation schemes without performance degradation.

\subsection{Decoupling Principle}
As established in the previous subsection, the decoupling of baseband processing and beamforming constitutes the core design philosophy of the MSA. This system-level decoupling originates from the independent manipulation of the magnitude and phase of the RC at the unit cell level.
o elucidate this principle mathematically, we express the reflected electric field  $ E^{\operatorname{o}}(\Omega,t)$ as the product of a time-varying component and a space-varying component, i.e., $E^{\operatorname{o}}(\Omega,t)=\bar{E}^{\operatorname{o}}(\Omega)\tilde{E}^{\operatorname{o}}(t)$. 
According to array signal processing theory and communication principle, the spatial energy distribution $\bar{E}^{\operatorname{o}}(\Omega)$ corresponds to the dual-domain representation of the surface phase distribution, while the time-varying component  $\tilde{E}^{\operatorname{o}}(t)$ represents the information-bearing signal that should be independent of the spatial direction $\Omega$. 
Based on the array reflection model in Eq.~\eqref{array reflection model}, and assuming that the time-varying component of the incident EM field is identical for all angles of arrival (AoAs) and that the time-varying magnitudes are uniform across all unit cells (i.e., $\tilde{\alpha}_k(t)=\tilde{\alpha}(t)$ for all $k$), the reflected electric field can be expressed as
\begin{equation}
    E^{\operatorname{o}}(\Omega,t)=\tilde{E}^{\operatorname{i}}(t)\tilde{\alpha}(t)\cdot\iint_{\mathcal{S}}\operatorname{d}\Omega^{'}\mathbf{w}^{\top}\mathbf{v}(\Omega,\Omega^{'})\bar{E}^{\operatorname{i}}(\Omega^{'}),
    \label{decoupling}
\end{equation}
where $\mathbf{w}=[e^{j\tilde{\beta}_1},\dots,e^{j\tilde{\beta}_K}]^{\top}$ denotes the phase steering vector of the metasurface, and $\mathbf{v}(\Omega,\Omega^{'})\in\mathbb{C}^{K}$ is the array manifold vector from incident direction $\Omega^{'}$ to reflection direction $\Omega$ with $k$-the element expressed as $\bar{\gamma}(\Omega,\Omega^{'})e^{-j(\mathbf{u}({\Omega})-\mathbf{u}({\Omega}^{'}))^{\top}\mathbf{p}_k}$. 

\begin{figure}[ht]
    \centering
        \includegraphics[width=0.7\linewidth]{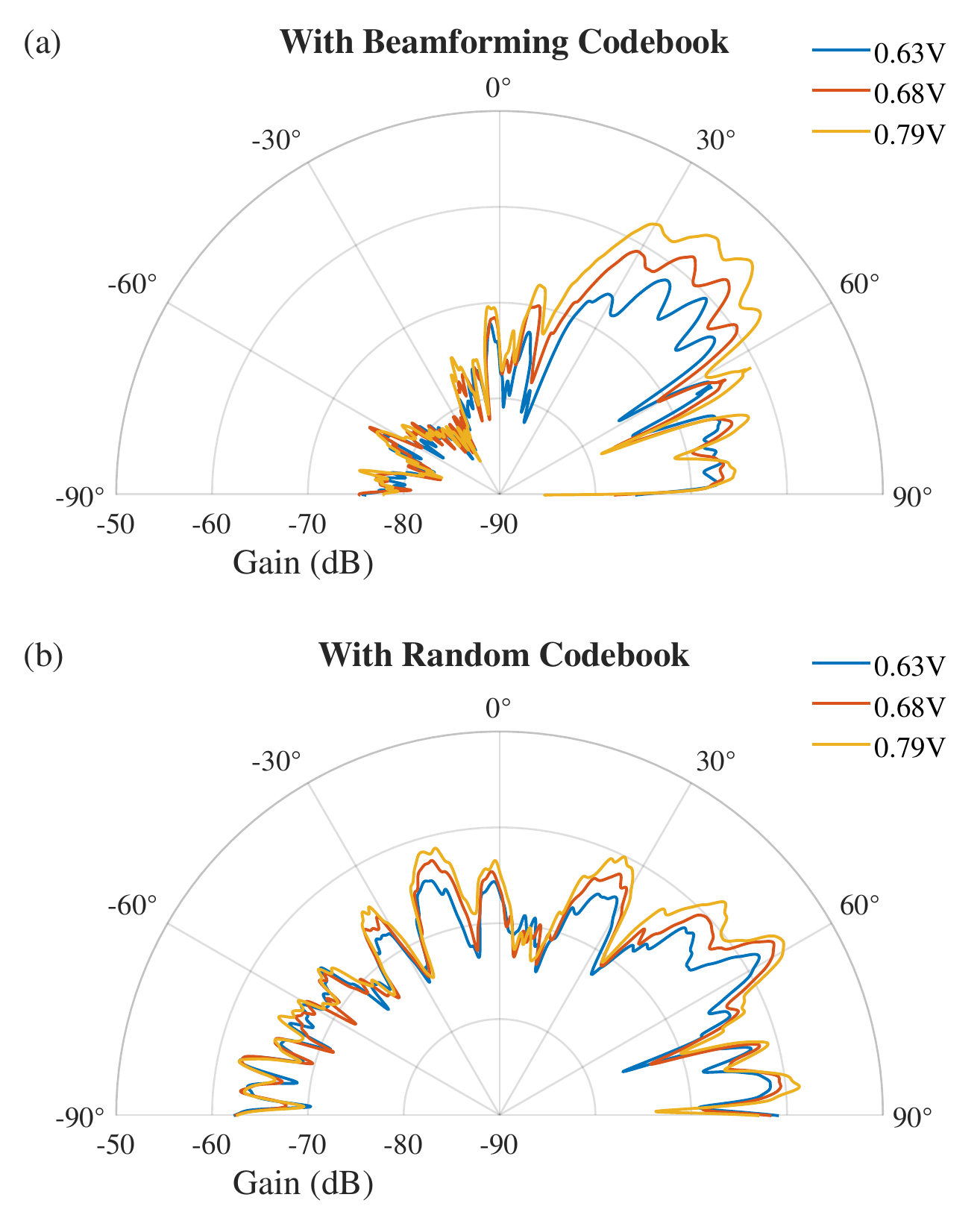}
        \caption{The measured radiation patterns in anechoic chamber under different bias voltage conditions while employing distinct codebook. (a) With beamforming codebook. (b) With random codebook.}
    \label{radiate patterns}
\end{figure}

According to Eq.~\eqref{decoupling}, it is evident that by independently controlling the magnitude $\tilde{\alpha}(t)$ and the phase distribution $\mathbf{w}$, we can achieve the desired decoupling. Specifically, the former one are designed to modulate the information-bearing signal $\tilde{E}^{\operatorname{o}}(t)$, while the latter one are configured to realize the spatial energy distribution $\bar{E}^{\operatorname{o}}(\Omega)$ for beam steering.
This separation ensures that information symbols remain consistent across different spatial locations, thereby eliminating symbol anisotropy. 
As shown in Fig.~\ref{radiate patterns}, the radiation patterns of the reflected EM fields remain nearly unchanged under different bias voltages (0.63 V, 0.68 V, and 0.79 V) corresponding to different RC magnitudes, which validates the effectiveness of the decoupling. 
Notably, when using the different codebooks, the main lobe direction and shape of the radiation pattern remain stable under varying voltages, indicating that the phase distribution dominates the beam steering while amplitude variation has negligible impact on the radiation patterns. It further highlights the robustness of the proposed decoupling approach.

\subsection{Superheterodyne Design}
To achieve decoupling, the modulation factors must remain independent of the phase factors—specifically, only the magnitude of the RC may vary. While this constraint might suggest that only amplitude modulation is feasible under the decoupling framework described in Eq.~\eqref{decoupling}, we overcome this apparent limitation by innovatively incorporating the superheterodyne concept into the MSA architecture.
Given one complex baseband symbol $x=a+jb\in\mathbb{C}$ where $a$ is the real part and the image part, the traditional zero-IF transmitter will directly up-converse it to RF by I/Q modulation as
\begin{equation}
    x_{\operatorname{RF}}(t)=[a\operatorname{cos}(2\pi f_{\operatorname{RF}}t)-b\operatorname{sin}(2\pi f_{\operatorname{RF}}t)]\left(u(t)-u(t-T_s)\right),
\end{equation}
where $f_{\operatorname{RF}}$ denotes the RF, $u(t)$ is the step function and $T_s$ is the symbol duration. Essentially, the $\pi/2$ phase difference between the real and imaginary parts of the carrier wave in the above equation is defined relative to the RF frequency. This direct conversion architecture presents significant challenges for metasurface-based transmitters attempting to independently control symbol phase and beamforming phase, due to the inherent hardware coupling.

To overcome this limitation while maintaining symbol isotropy, we introduce an iIF stage, inspired by conventional superheterodyne transmitter architectures as illustrated in the DUC module of Fig.~\ref{Procedure}(b). The complex baseband symbol is first upconverted to IF
\begin{equation}
    x_{\operatorname{IF}}(t)=[a\operatorname{cos}(2\pi f_{\operatorname{IF}}t)-b\operatorname{sin}(2\pi f_{\operatorname{IF}}t)]\left(u(t)-u(t-T_s)\right),
    \label{DUC}
\end{equation}
where $f_{\operatorname{IF}}\ll f_{\operatorname{RF}}$.This DUC process, described by Eq.~\eqref{DUC}, can be efficiently implemented and converted to an analog IF signal $x_{\operatorname{IF}}(t)$ using a DAC. 
The resulting IF signal is upconverted to RF through mixing with the incident carrier signal via the metasurface-based multiplier as
\begin{equation}
    x_{\operatorname{RF}}(t)=x_{\operatorname{IF}}(t)\cdot\operatorname{cos(2\pi f_{RF}t)}e^{j\tilde{\beta}},
    \label{}
\end{equation}
where $x_{\operatorname{RF}}(t)$ denotes the reflected signal and the $e^{j\tilde{\beta}}$ denotes the phase factor. By introducing the IF stage, the phase of the transmitted symbols becomes decoupled from the RF carrier phase, enabling complete separation between complex signal generation and beamforming operations. This architecture ensures that identical complex symbols are received consistently across all spatial locations.

A key advantage of the MSA architecture is its inherent scalability to multi-stream transmission. The number of independent data streams that the MSA transmitter can support is determined by the number of DUC modules integrated into the system. Each DUC module can generate an independent baseband (or IF) signal stream, denoted as $x_{m}(t)$ for the $m$-th stream. These streams can be simultaneously fed into the MPD-metasurface, which modulates and reflects each stream onto the incident RF carrier. Spatial multiplexing of these streams is achieved through a hierarchical precoding strategy, as detailed in Section~\ref{multi streams}.

\section{Modeling of MSA\label{Modeling of MSA}}
This section develops a hierarchical modeling framework for the MSA, beginning with a unit-level reflection model that characterizes the spatial and temporal dynamics of individual meta-atoms. It then extends to an array-level analysis incorporating phase differences and transformation matrices for collective scattering behavior. Finally, it establishes a comprehensive system-level signal model for bi-static MIMO communications.
\subsection{Unit Reflection Model}
Denote the spatial direction by $\Omega\triangleq(\theta,\phi)$, where $\theta$ and $\phi$ are the zenith and azimuth angles, respectively.

\begin{equation}
    \begin{aligned}
    \Gamma(\Omega_{\operatorname{o}},\Omega_{\operatorname{i}},t)=\alpha(\Omega_{\operatorname{o}},\Omega_{\operatorname{i}},t)e^{j\psi(\Omega_{\operatorname{o}},\Omega_{\operatorname{i}},t)},
    \end{aligned}
\end{equation}
where the RCs $\Gamma(\Omega_{\operatorname{o}},\Omega_{\operatorname{i}},t)\in\mathbb{C}$ represents the ratio of the scattered electrical field to the incident electrical field for given incident direction $\Omega_{\operatorname{i}}$ and reflection direction $\Omega_{\operatorname{o}}$ at time instant $t$. The RC can also be expressed in terms of the unit pattern component and the reconfigurable component as
\begin{equation}
    \Gamma(\Omega_{\operatorname{o}},\Omega_{\operatorname{i}},t)=\bar{\Gamma}(\Omega_{\operatorname{o}},\Omega_{\operatorname{i}})\tilde{\Gamma}(t).
\end{equation}
In this model, the overall response is governed by two part: $\bar{\Gamma}(\Omega_{\operatorname{o}},\Omega_{\operatorname{i}})\in\mathbb{C}$ describes the fixed spatial scattering property of the unit for given angles, while $\tilde{\Gamma}(t)\in\mathbb{C}$ embodies the active time-varying reconfigurable component. Based on the reciprocity theorem~\cite{balanis2016antenna}, the angular response of a reciprocal meta-atom is often separable and can be modeled as the product of a single element pattern function $F(\Omega) $ evaluated for the incident and reflection directions:
\begin{equation}
    \bar{\Gamma}(\Omega_i, \Omega_o) = F(\Omega_i) F(\Omega_o),
\end{equation}
\begin{equation}
    \tilde{\Gamma}(t)=\alpha(t)e^{j\beta(t)},
\end{equation}
where $\alpha(t)$ and $\beta(t)$ respectively denotes the time-varying magnitude and phase of RC. The scattering electrical fields of one unit can be expressed as
\begin{equation}
    E^{\operatorname{o}}(\Omega,t)=\tilde{\Gamma}(t)\iint_{\mathcal{S}}E^{\operatorname{i}}(\Omega^{'},t)\bar{\Gamma}(\Omega,\Omega^{'})\operatorname{d}\Omega^{'},
    \label{continuous}
\end{equation}
where the integral is taken over the domain $\mathcal{S}$, which represents the entire front half-sphere of incident directions. This assumes the metasurface lies in the plane $z=0$ and $\mathcal{S} = \{ \Omega' = (\theta', \phi') \mid 0 \le \theta' < \pi/2, 0 \le \phi' < 2\pi \}$.

To facilitate a more concise representation of the relationship between the metasurface and the wireless channel at both ends, we discretize the spatial angles and reformulate the continuous model in Eq~\eqref{continuous} into a matrix multiplication form.
We discretize the angular hemisphere $\mathcal{S}$  into $M$ directions $\{\Omega_1, \Omega_2, \dots, \Omega_M\}$. At time instant $t$, we define the incident/outgoing field vector as
\begin{equation}
    \mathbf{e}^{\operatorname{i/o}}(t)=[E^{\operatorname{i/o}}(\Omega_1,t),\dots,E^{\operatorname{i/o}}(\Omega_M,t)]^\top,
\end{equation}
where $\mathbf{e}^{\operatorname{i/o}}\in \mathbb{C}^{M}$, whose $m$-th element represents the complex incident/outgoing field of the $m$-th direction $\Omega_m$. The pattern function is discretized in angular domain as 
\begin{equation}
\mathbf{f}=[F(\Omega_1),\dots,F(\Omega_m)]^{\top},
\end{equation}
where $\mathbf{f}\in\mathbb{C}^{M}$, then the discretized version of scattering electrical fields in~\eqref{continuous} can be expressed as
\begin{equation}
    \mathbf{e}^{\operatorname{o}}(t)=\tilde{\Gamma}(t)\cdot\mathbf{f}\mathbf{f}^\dagger\mathbf{e}^{\operatorname{i}}(t).
\end{equation}

\subsection{Array Reflection Model}
Given the far-filed assumption, the incident directional vector $\mathbf{u}(\Omega)$ can be expressed as
\begin{equation}
    \begin{aligned}
        \mathbf{u}(\Omega)=
        \left[\begin{array}{l}
        \sin \theta \cos \phi \\
        \sin \theta \sin \phi \\
        \cos \theta
        \end{array}\right]
    \end{aligned},
\end{equation}
and the equivalent electrical position vector of $k$-th unit's center at $i$-th row and $j$-th column is $\mathbf{p}_{k}=\frac{2\pi\Delta d}{\lambda}\cdot[i,j,0]^{\operatorname{T}}$, where $k=(i-1)*K_r+j$ and $\lambda$ is the wavelength of the center frequency. The phase difference of the incident wave and outgoing wave at the $k$-th element relative to the origin can be expressed as $e^{-j\mathbf{u}(\Omega)^\top\mathbf{p}_k}$. We define the phase difference matrix of far fields as
\begin{equation}
    \mathbf{U}=\left[\begin{array}{ccc}
                e^{-j\mathbf{u}(\Omega_1)^\top\mathbf{p}_1} & \cdots & e^{-j\mathbf{u}(\Omega_M)^\top\mathbf{p}_1} \\
                \vdots & & \vdots \\
                e^{-j\mathbf{u}(\Omega_1)^\top\mathbf{p}_K} & \cdots & e^{-j\mathbf{u}(\Omega_M)^\top\mathbf{p}_K}
                \end{array}\right].
\end{equation}

The scattered electric field vector in the angular domain for the metasurface can be expressed as
\begin{equation}
    \mathbf{e}^\text{o}(t) = \mathbf{W}^{\dagger} \mathbf{\tilde{\Gamma}}(t) \mathbf{W} \mathbf{e}^\text{i}(t),
    \label{array reflection model}
\end{equation}
where $\mathbf{W} = \mathbf{U} \operatorname{diag}\{\mathbf{f}\} \in \mathbb{C}^{K \times M}$ denotes the static component of the transformation from the incident electric field to the scattered field, and $\mathbf{\tilde{\Gamma}}(t) = \operatorname{diag}\{[\tilde{\Gamma}_1(t), \tilde{\Gamma}_2(t), \dots, \tilde{\Gamma}_K(t)]^\top\} \in \mathbb{C}^{K \times K}$ represents the reconfigurable part of the transformation. 
Equation~\eqref{array reflection model} describes the overall scattering behavior of the metasurface array. Specifically, the incident field $\mathbf{e}^\text{i}(t)$ is first transformed into the element domain via $\mathbf{W}$. This operation, corresponding to left-multiplication by $\mathbf{W}$, implements the coherent superposition of the incident time-varying electric fields from all angles of arrival (AoAs) at each unit. The result is then modulated by the diagonal matrix $\mathbf{\tilde{\Gamma}}(t)$, which encapsulates the reconfigurable RCs of individual units. Finally, the signal is transformed back to the angular domain through $\mathbf{W}^{\dagger}$ to produce the scattered field $\mathbf{e}^\text{o}(t)$.

In our proposed MSA, each unit cell is designed to independently control both the magnitude and phase of the RC. Specifically, the magnitude carries the information symbol, while the phase is adjusted to achieve passive beamforming towards desired directions. This decoupled control mechanism allows for flexible modulation schemes and enhanced spatial manipulation of the backscattered signals. Denoteing the reconfigurable part of the transformation matrix as the multiplication of magnitude and phase components, during a coherent time, we have
\begin{equation}
    \mathbf{e}^\text{o}(t) = \mathbf{W}^{\dagger} \mathbf{\Lambda}(t)\mathbf{\Phi} \mathbf{W} \mathbf{e}^\text{i}(t),
    \label{array reflection model2}
\end{equation}
where $\mathbf{\Lambda}(t) = \operatorname{diag}\{[\alpha_1(t), \alpha_2(t), \dots, \alpha_K(t)]^\top\} \in \mathbb{C}^{K \times K}$ represents the magnitude modulation matrix, and $\mathbf{\Phi} = \operatorname{diag}\{[e^{j\phi_1}, e^{j\phi_2}, \dots, e^{j\phi_K}]^\top\} \in \mathbb{C}^{K \times K}$ denotes the phase adjustment matrix for beamforming.

\subsection{Overall Signal Model}
Consider a MSA assisted bistatic MIMO system consisting of a transmitting base station (Tx), a receiving base station (Rx), and a MSA backscatter transmitter. Both BSs are equipped with uniform linear arrays (ULAs) of $N_t$ and $N_r$ antennas, respectively, while the MSA comprises $K$ programmable unit cells arranged in a rectangular grid with $K_r$ rows and $K_c$ columns ($K=K_r\times K_c$). The inter-element spacing of the MSA is denoted by $\Delta d$.

The transmitting BS emits an RF signal that propagates through the wireless channel to the MSA, which modulates its RCs to encode information onto the backscattered signal. 
This modulated signal then traverses another wireless channel to reach the receiving BS. In order to establish a comprehensive system model considering the symbol-level spatial isotropy, we model the wireless channels from antenna domain of the Tx to angular domain of the MSA and from angular domain of the MSA to antenna domain of the Rx as $\mathbf{H}_{\text{tx-MSA}}\in\mathbb{C}^{M\times N_t}$ and $\mathbf{H}_{\text{MSA-rx}}\in\mathbb{C}^{N_r\times M}$, respectively. 
The channels are assumed to be quasi-static flat-fading, remaining constant over a transmission block. The transmitted signal vector from the Tx is denoted as $\mathbf{x}(t)\in\mathbb{C}^{N_t}$, and the received signal vector $\mathbf{y}(t)\in\mathbb{C}^{N_r}$ at the Rx is represented as
\begin{equation}
    \mathbf{y}(t) = \mathbf{H}_{\text{MSA-rx}} \mathbf{W}^{\dagger} \mathbf{\Lambda}(t) \mathbf{\Phi} \mathbf{W} \mathbf{H}_{\text{tx-MSA}} \mathbf{x}(t) + \mathbf{n}(t),
    \label{signal model}
\end{equation}
where \(\mathbf{x}(t) \in \mathbb{C}^{N_t}\) is the transmitted signal, \(\mathbf{n}(t) \in \mathbb{C}^{N_r}\) is additional white Gaussian noise with covariance \(\sigma^2 \mathbf{I}_{N_r}\).
The channel matrices are expressed using multipath decompositions. The MSA-to-Rx channel is:
\begin{equation}
    \mathbf{H}_{\text{MSA-rx}} = \sum_{l=1}^{L} \alpha_l e^{-j2\pi f_c \tau_l} \mathbf{a}_{\text{r}}(\Omega_l^{\text{r}}) \mathbf{v}(\Omega_l^{\text{o}})^{\top},
\end{equation}
where \(\mathbf{a}_{\text{r}}(\Omega_l^{\text{r}}) = [e^{-j\mathbf{u}(\Omega^{\text{r}}_l)^{\top} \mathbf{p}^{\text{r}}_1}, \dots, e^{-j\mathbf{u}(\Omega^{\text{r}}_l)^{\top} \mathbf{p}^{\text{r}}_{N_r}}]^{\top}\) is the Rx array response vector, and \(\mathbf{v}(\Omega_l^{\text{o}}) = [\delta(\Omega_1-\Omega_l^{\text{o}}), \dots, \delta(\Omega_M-\Omega_l^{\text{o}})]^{\top}\) is the angular selection vector (with \(\delta(\cdot)\) approximatable by sinc/Dirichlet functions in practice). Similarly, the Tx-to-MSA channel is:
\begin{equation}
    \mathbf{H}_{\text{tx-MSA}} = \sum_{q=1}^{Q} \beta_q e^{-j2\pi f_c \zeta_q} \mathbf{v}(\Omega_q^{\text{i}}) \mathbf{a}_{\text{t}}(\Omega_q^{\text{t}})^{\top}.
\end{equation}

Defining the simplified channel matrices \(\mathbf{H}_{\text{i}} = \mathbf{W} \mathbf{H}_{\text{tx-MSA}}\) and \(\mathbf{H}_{\text{o}} = \mathbf{H}_{\text{MSA-rx}} \mathbf{W}^{\dagger}\), which expand to:
\begin{equation}
\mathbf{H}_{\text{i}} = \sum_{q=1}^Q \beta_q e^{-j2\pi f_c \zeta_q} F(\Omega_q^{\text{i}}) \mathbf{a}_{\text{MSA}}(\Omega_q^{\text{i}}) \mathbf{a}_{\text{t}}(\Omega_q^{\text{t}})^{\top},
\end{equation}
\begin{equation}
    \mathbf{H}_{\text{o}} = \sum_{l=1}^L \alpha_l e^{-j2\pi f_c \tau_l} F^{\dagger}(\Omega_l^{\text{o}}) \mathbf{a}_{\text{r}}(\Omega_l^{\text{r}}) \mathbf{a}_{\text{MSA}}(\Omega_l^{\text{o}})^{\dagger},
\end{equation}
where \(\mathbf{a}_{\text{MSA}}(\Omega) = [e^{-j \mathbf{u}(\Omega)^{\top} \mathbf{p}_1}, \dots, e^{-j \mathbf{u}(\Omega)^{\top} \mathbf{p}_K}]^{\top}\), the signal model simplifies to:
\begin{equation}
    \mathbf{y}(t) = \mathbf{H}_{\text{o}} \mathbf{\Lambda}(t) \mathbf{\Phi} \mathbf{H}_{\text{i}} \mathbf{x}(t) + \mathbf{n}(t).
    \label{final signal model}
\end{equation}
This formulation provides a concise expression of the end-to-end transmission while maintaining all essential components.

\section{Functional Analysis\label{Functional Analysis}}
This section provides a comprehensive analysis of the core performance advantages of the MSA architecture, based on the proposed signal model. 
To highlight the essential characteristics of the MSA, we adopt several reasonable simplifications in our analysis. We assume that the incident transmission signal is an ideal sinusoidal carrier wave, expressed as $\mathbf{x}(t)=\mathbf{w}_ts_c$, where $s_c$ is a complex constant representing the constant envelope of carrier and $\mathbf{w}_t$ is the transmitting beamforming vector. Although the use of modulated ambient waves is a relevant consideration in backscatter communication scenarios, such configurations are beyond the scope of this work. 

\subsection{Symbol-level Isotropy}
Symbol-level isotropy represents a fundamental performance metric for metasurface-based transmitters, characterizing the consistency of received symbols across different spatial locations. 
To achieve high-order QAM modulation using low-bit metasurfaces, advanced codebook design techniques such as space-time coding schemes are typically employed to approximate desired constellation points~\cite{tang2020wireless,zhang2021wireless}. However, these methods often introduce symbol anisotropy, where different spatial locations receive varying symbol constellations due to the inherent coupling between modulation and beamforming processes. While directional symbol variations may be desirable in specific applications like physical-layer security, uniform symbol reception is generally preferred in most wireless communication systems.

Symbol anisotropy can significantly degrade communication performance and limit system applicability, particularly in multipath or time-varying environments. In multipath scenarios, signals reflected from different spatial directions experience varying losses and delays before reaching the receiver. 
For traditional transmitters with omnidirectional symbol transmission, receivers can coherently combine these multipath components to effectively utilize the diversity and recover the original symbols. 
However, for conventional metasurface-based transmitters exhibiting symbol anisotropy, the reception of different constellations from different paths complicates the decoding process and increases error rates.

The coexistence of high-order QAM transmission and symbol-level isotropy constitutes a distinctive advantage of the MSA architecture.  The essential reason to achieve this advantage lies in the following two aspects: (1) the signal generated by individual unit cells should be identical in terms of symbol; (2) the beamforming phase distribution should not affect the transmitted symbols.
By decoupling the baseband processing from beamforming, the MSA ensures that all spatial directions receive identical symbol constellations, regardless of the beam steering configuration. Assuming the generated IF signals imposed into each unit are identical, the reflected electric field vector can be rewritten as
\begin{equation}
    \mathbf{e}^\text{o}(t) = \alpha(t) \mathbf{W}^{\dagger} \mathbf{\Phi} \mathbf{W} \mathbf{H}_{\text{tx-MSA}} \mathbf{w}_ts_c,
    \label{isotropy}
\end{equation}
where each element of $\mathbf{e}^\text{o}(t)$ shares the same $\alpha(t)$, ensuring symbol isotropy across all spatial directions.

\subsection{Arbitrary Waveform Generation without Harmonic}
The modular and decoupled design of the MSA enables arbitrary waveform generation with significantly higher fidelity than conventional metasurface-based transmitters. This capability stems from a fundamental architectural innovation: signal generation is delegated to a dedicated DUC module, while the MPD-metasurface operates within the highly linear region of its reflection characteristics, facilitating high-fidelity, low-distortion waveform synthesis.

\begin{figure}[ht]
    \centering
        \includegraphics[width=\linewidth]{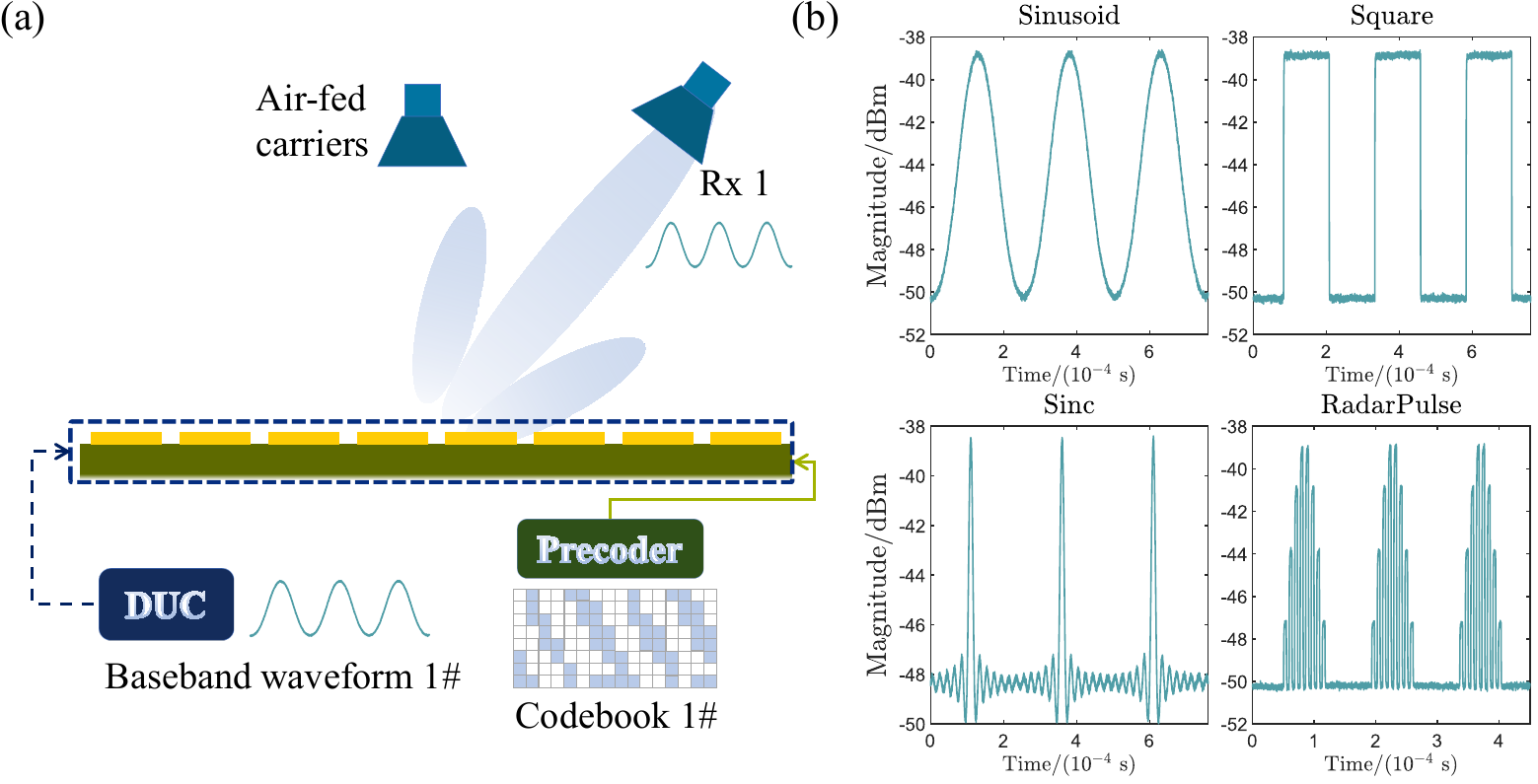}
        \caption{(a) The illustration of arbitrary waveform generation setup utilizing single DUC module. (b) The experimental results of various fundamental waveforms generated by MSA.}
    \label{arbitrary waveform}
\end{figure}

Conventional metasurfaces relying on nonlinear switching mechanisms approximate target waveforms through rapid diode state toggling, which introduces inherent nonlinearity that causes severe harmonic interference and signal distortion. The MSA fundamentally overturns this paradigm. As shown in Fig.~\ref{arbitrary waveform}(a), the DUC module performs precise baseband signal processing in the digital domain, producing a pristine baseband analog voltage signal that corresponds to the desired waveform. Subsequently, the MPD-metasurface functions as a reconfigurable reflective modulator and beamformer, whose primary role is to linearly modulate the amplitude of the incident RF carrier according to the input analog voltage signal and to manipulate the beampatterns. 

The flexibility of the DUC module enables the synthesis of arbitrarily analog waveforms, ranging from high-order QAM signals for communication to sophisticated radar waveforms for sensing applications (see Fig.~\ref{MSA-Backscatter} and Fig.~\ref{MSA-Doppler-Spoofing}). The linear response of the MPD-metasurface ensures these waveforms are reproduced with high fidelity in the RF domain. Experimental results in Fig.~\ref{arbitrary waveform}(b) verify that the MSA can accurately generate and superpose various fundamental waveforms (e.g., sinusoidal and etc.) without any distortion.
\begin{figure}[ht]
    \centering
        \includegraphics[width=0.9\linewidth]{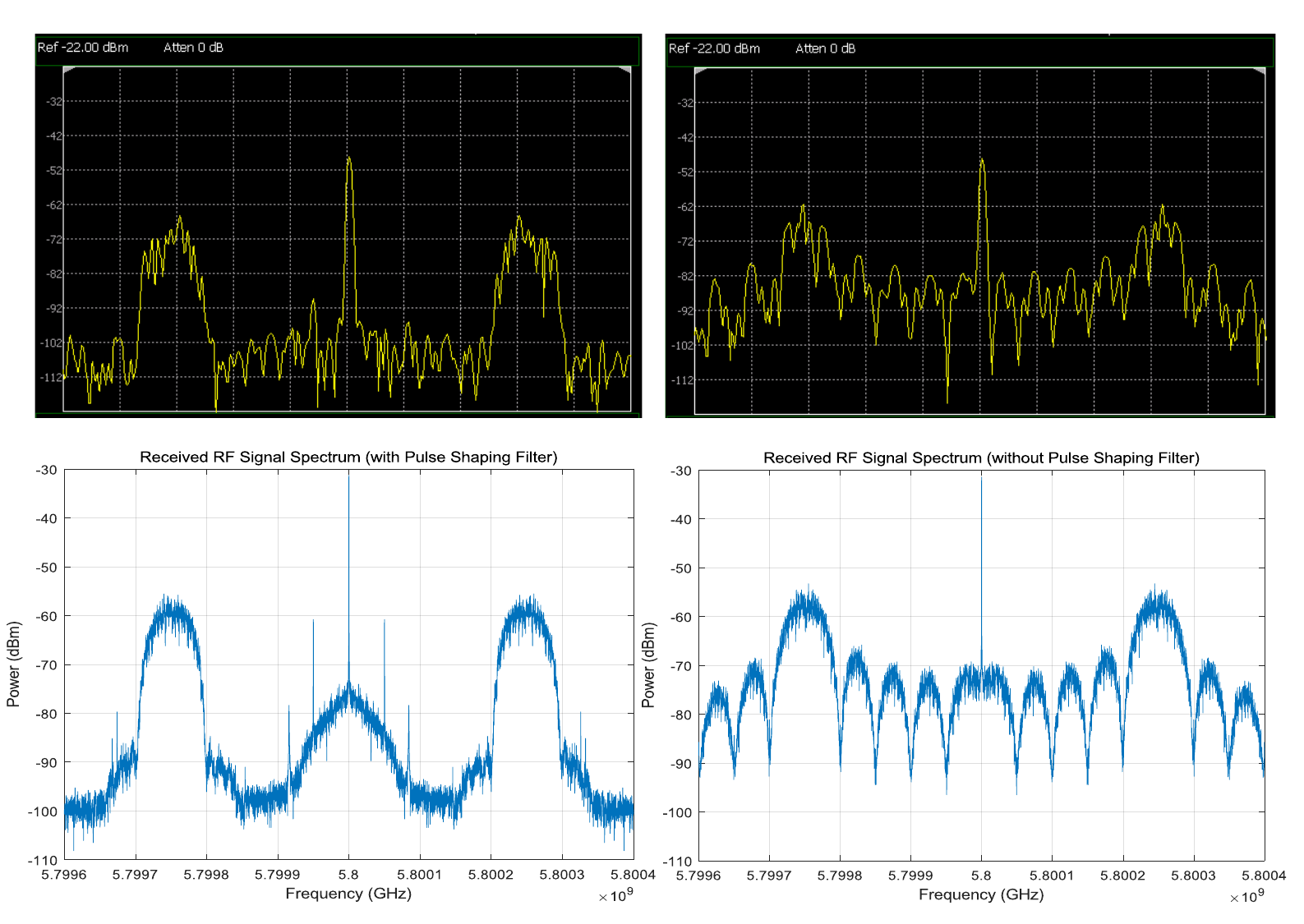}
        \caption{Screenshots of the handheld spectrum analyzer and the received signal spectrum diagrams by the USRP for a 256-QAM single-carrier signal generated by MSA with and without raised-cosine pulse shaping filter}
    \label{harmonic}
\end{figure}

By getting rid of the nonlinear switching process, the MSA significantly mitigates harmonic interference.
By avoiding abrupt diode switching, the MPD-metasurface produces a reflected spectrum that is dominated by the intended modulation sidebands while suppressing spurious harmonic components. When combined with appropriate pulse‑shaping and filtering in the DUC chain, spectral sidelobes are further reduced, yielding very low out‑of‑band leakage. As illustrated in Fig.~\ref{harmonic}(a), we compared the spectrum of a 256-QAM single-carrier signal generated without and with raised-cosine pulse shaping filters. The results demonstrate that the MSA architecture has the ability to eliminate the spurious emissions that are typically present in conventional metasurface-based transmitters.

The capability to generate arbitrary waveforms without harmonic interference establishes the foundation for multi-functional applications within the MSA architecture, such as simultaneous high-quality communications and high-fidelity sensing. This ensures high spectral efficiency for communication signals while maintaining waveform authenticity and reliability for applications like radar spoofing.

\subsection{Diversity Gain via Passive Precoding\label{diversity section}}
In this section, we derive the closed-form solution for optimizing the phase matrix \(\mathbf{\Phi}\) in the MSA architecture to maximize the received signal-to-noise ratio (SNR), and analyze the relationship between diversity gain and the number of unit cells \(K\). This analysis highlights the fundamental advantages of MSA in harnessing spatial degrees of freedom for enhanced reliability.
The received signal model is 
\begin{equation}
\mathbf{y}(t) = \alpha(t) \mathbf{H}_{\text{o}} \mathbf{\Phi} \mathbf{H}_{\text{i}} \mathbf{w}_t s_c + \mathbf{n}(t),
\end{equation}
where \(\mathbf{\Phi} = \operatorname{diag}[e^{j\phi_1}, e^{j\phi_2}, \dots, e^{j\phi_K}]\) is the phase adjustment matrix. The goal is to maximize the SNR, which is equivalent to maximize the power of the desired signal component \(\| \mathbf{H}_{\text{o}} \mathbf{\Phi} \mathbf{H}_{\text{i}} \mathbf{w}_t \|^2\). Define the effective channel vector as \(\mathbf{h}_{\text{eff}} = \mathbf{H}_{\text{i}} \mathbf{w}_t \in \mathbb{C}^{K \times 1}\), the optimization problem reduces to
\begin{equation}
\max_{\mathbf{\Phi}} \| \mathbf{H}_{\text{o}} \mathbf{\Phi} \mathbf{h}_{\text{eff}} \|^2 \quad \text{subject to} \quad |[\mathbf{\Phi}]_{ii}| = 1 \ \forall i.
\end{equation}
To solve this, we perform a singular value decomposition of \(\mathbf{H}_{\text{o}} = \mathbf{U} \boldsymbol{\Sigma} \mathbf{V}^H\), where \(\mathbf{V} = [\mathbf{v}_1, \mathbf{v}_2, \dots, \mathbf{v}_K]\) contains the right singular vectors, and \(\mathbf{v}_1\) corresponds to the largest singular value \(\sigma_1\). The optimal phase alignment is achieved when the combined vector \(\mathbf{\Phi} \mathbf{h}_{\text{eff}}\) is phase-matched to \(\mathbf{v}_1\). This yields the closed-form solution for each phase angle:
\begin{equation}
\phi_i = \arg(v_{1,i}) - \arg([\mathbf{h}_{\text{eff}}]_i), \quad i = 1, 2, \dots, K,
\end{equation}
where \(v_{1,i}\) is the \(i\)-th element of \(\mathbf{v}_1\), and \([\mathbf{h}_{\text{eff}}]_i\) is the \(i\)-th element of \(\mathbf{h}_{\text{eff}}\). The optimal phase matrix is then \(\mathbf{\Phi}_{\text{opt}} = \operatorname{diag}[e^{j\phi_1}, e^{j\phi_2}, \dots, e^{j\phi_K}]\).
This solution ensures coherent combining of the signal paths through the dominant mode of \(\mathbf{H}_{\text{o}}\), maximizing the received power. 

Diversity gain quantifies the improvement in link reliability by exploiting independent fading paths. In the MSA, the diversity order is determined by the number of effectively uncorrelated channels introduced by the \(K\) unit cells. The maximal diversity order achievable is bounded by the number of independent paths in the channel, which scales with \(K\) when the unit cells are spaced sufficiently apart to ensure low correlation.
The average received power with optimal phase optimization can be expressed as:
\begin{equation}
\mathbb{E}[\| \mathbf{H}_{\text{o}} \mathbf{\Phi}_{\text{opt}} \mathbf{h}_{\text{eff}} \|^2] = \sigma_1^2 \| \mathbf{h}_{\text{eff}} \|^2,
\end{equation}
where \(\sigma_1\) is the largest singular value of \(\mathbf{H}_{\text{o}}\). For a rich scattering environment, the singular values of \(\mathbf{H}_{\text{o}}\) scale with \(\sqrt{K}\) due to the increased spatial degrees of freedom. Consequently, the received power grows linearly with \(K\), i.e., \(\mathbb{E}[\| \mathbf{H}_{\text{o}} \mathbf{\Phi}_{\text{opt}} \mathbf{h}_{\text{eff}} \|^2] \propto K\). This translates to a diversity order of \(K\) in the high-SNR regime, as the error probability decays as \(P_e \propto \text{SNR}^{-K}\).

However, the actual diversity gain depends on the correlation between the unit cells. If the unit spacing is below the coherence distance, the gain saturates due to correlated fading. For a half-wavelength spacing (\(\Delta d = \lambda/2\)), the diversity order approaches \(K\) asymptotically. This relationship is validated through simulations, showing that the bit error rate (BER) improves significantly with increasing \(K\), particularly in multipath environments.

\subsection{Multi-stream transmission Design\label{multi streams}}
In this subsection, we analyze the multi-stream transmission capability of the MSA architecture using the proposed signal model. We consider a two-stream scenario to illustrate the underlying principles, which can be extended to more streams straightforwardly.
\begin{figure}[ht]
    \centering
        \includegraphics[width=\linewidth]{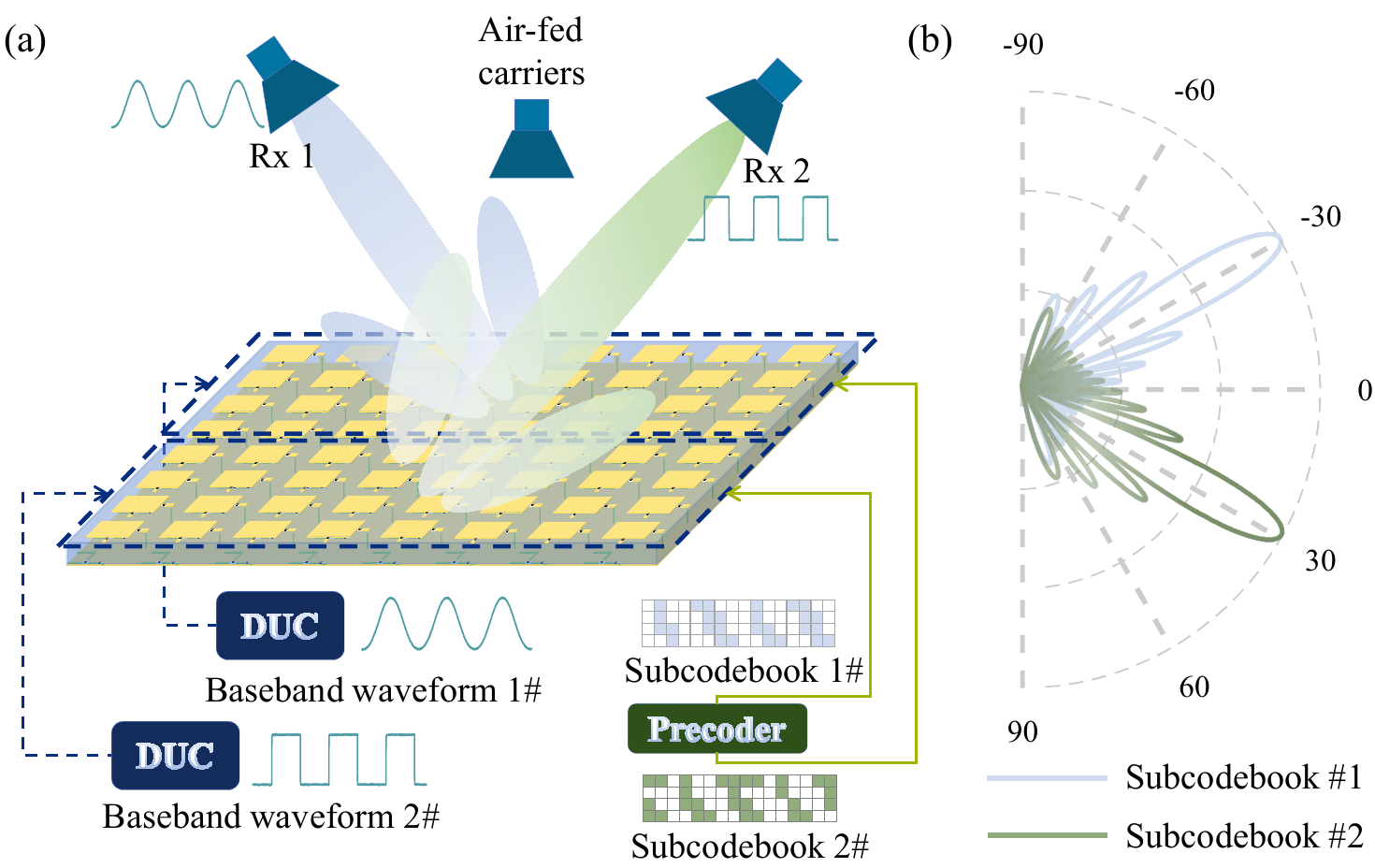}
        \caption{(a). illustration of two-stream MSA transmitter. (b). Schematic plot of the precoding patterns of two streams after joint phase optimization.}
    \label{two streams}
\end{figure}
We partition the MSA into two sub-surfaces, each part equipped with one DUC module and responsible for transmitting one data stream. The received signal can be expressed as:
\begin{equation}
\begin{aligned}
        &\mathbf{y}(t)\\
        =&\left[\begin{array}{ll}
        \mathbf{H}_{\operatorname{o}}^{11}&\mathbf{H}_{\operatorname{o}}^{12} \\
        \mathbf{H}_{\operatorname{o}}^{21}&\mathbf{H}_{\operatorname{o}}^{22}
        \end{array}\right]\left[\begin{array}{ll}
        \mathbf{\Lambda}_{1}(t)\mathbf{\Phi}_{1}& \\
        &\mathbf{\Lambda}_{2}(t)\mathbf{\Phi}_{2}
        \end{array}\right]\mathbf{h}_{\text{eff}}s_c+\mathbf{n}(t)\\
        =&\left[\begin{array}{ll}
        \mathbf{H}_{\operatorname{o}}^{11}\mathbf{\Lambda}_{1}(t)\mathbf{\Phi}_{1}+\mathbf{H}_{\operatorname{o}}^{12}\mathbf{\Lambda}_{2}(t)\mathbf{\Phi}_{2} \\
        \mathbf{H}_{\operatorname{o}}^{21}\mathbf{\Lambda}_{1}(t)\mathbf{\Phi}_{1}+\mathbf{H}_{\operatorname{o}}^{22}\mathbf{\Lambda}_{2}(t)\mathbf{\Phi}_{2}
        \end{array}\right]\mathbf{h}_{\text{eff}}s_c+\mathbf{n}(t)\\
        =&\left[\begin{array}{ll}
        \alpha_{1}(t)\mathbf{H}_{\operatorname{o}}^{11}\mathbf{\Phi}_{1}\mathbf{h}_{\text{eff}}+\alpha_{2}(t)\mathbf{H}_{\operatorname{o}}^{12}\mathbf{\Phi}_{2}\mathbf{h}_{\text{eff}} \\
        \alpha_{1}(t)\mathbf{H}_{\operatorname{o}}^{21}\mathbf{\Phi}_{1}\mathbf{h}_{\text{eff}}+\alpha_{2}(t)\mathbf{H}_{\operatorname{o}}^{22}\mathbf{\Phi}_{2}\mathbf{h}_{\text{eff}}
        \end{array}\right]s_c+\mathbf{n}(t),
\end{aligned}
\end{equation}
where \(\mathbf{H}_{\operatorname{o}}^{ij}\) represents the channel from the \(j\)-th sub-surface to the \(i\)-th receiver, \(\mathbf{\Lambda}_{m}(t)=\alpha_m(t)\mathbf{I}\) and \(\mathbf{\Phi}_{m}=\operatorname{diag}[e^{j\phi_{(m-1)K/2+1}},\dots,e^{j\phi_{mK/2}}]\) for \(m=1,2\). The effective channel vector \(\mathbf{h}_{\text{eff}}\) is assumed to be identical for both sub-surfaces for simplicity. The signal-to-interference-plus-noise ratio (SINR) at each receiver can be expressed as:
\begin{equation}
    \operatorname{SINR}_1=\frac{\vert\alpha_{1}(t)\vert^2\vert\mathbf{H}_{\operatorname{o}}^{11}\mathbf{\Phi}_{1}\mathbf{h}_{\text{eff}}\vert^2}{\vert\alpha_{2}(t)\vert^2\vert\mathbf{H}_{\operatorname{o}}^{12}\mathbf{\Phi}_{2}\mathbf{h}_{\text{eff}}\vert^2+\sigma^2},
\end{equation}
\begin{equation}
    \operatorname{SINR}_2=\frac{\vert\alpha_{2}(t)\vert^2\vert\mathbf{H}_{\operatorname{o}}^{22}\mathbf{\Phi}_{2}\mathbf{h}_{\text{eff}}\vert^2}{\vert\alpha_{1}(t)\vert^2\vert\mathbf{H}_{\operatorname{o}}^{21}\mathbf{\Phi}_{1}\mathbf{h}_{\text{eff}}\vert^2+\sigma^2},
\end{equation}
where the magnitude factors $\alpha(t)$ are simplified as $1$ for notation convenience. The objective is to jointly optimize the phase matrices \(\mathbf{\Phi}_{1}\) and \(\mathbf{\Phi}_{2}\) to maximize the sum SINR, defined as \(\operatorname{SINR}_{\operatorname{sum}} = \operatorname{SINR}_1 + \operatorname{SINR}_2\):
\begin{equation}
    \operatorname{SINR}_{\operatorname{sum}}=\frac{\vert\mathbf{b}_1^{T}\boldsymbol{\phi}_1\vert^2}{\vert\mathbf{b}_2^{T}\boldsymbol{\phi}_2\vert^2+\sigma^2}+\frac{\vert\mathbf{c}_2^{T}\boldsymbol{\phi}_2\vert^2}{\vert\mathbf{c}_1^{T}\boldsymbol{\phi}_1\vert^2+\sigma^2},
\end{equation}
where $\mathbf{b}_1^{T}=\mathbf{H}_{\operatorname{o}}^{11}\operatorname{diag}(\mathbf{h}_{\text{eff}})$, $\mathbf{b}_2^{T}=\mathbf{H}_{\operatorname{o}}^{12}\operatorname{diag}(\mathbf{h}_{\text{eff}})$, $\mathbf{c}_2^{T}=\mathbf{H}_{\operatorname{o}}^{22}\operatorname{diag}(\mathbf{h}_{\text{eff}})$ and $\mathbf{c}_1^{T}=\mathbf{H}_{\operatorname{o}}^{21}\operatorname{diag}(\mathbf{h}_{\text{eff}})$. The phase vectors are defined as \(\boldsymbol{\phi}_1 = [e^{j\phi_1}, e^{j\phi_2}, \dots, e^{j\phi_{N/2}}]^T\) and \(\boldsymbol{\phi}_2 = [e^{j\phi_{N/2+1}}, e^{j\phi_{N/2+2}}, \dots, e^{j\phi_N}]^T\). The optimization problem can be formulated as:
\begin{equation}
    \begin{aligned}
        \max_{\boldsymbol{\phi}_1,\boldsymbol{\phi}_2}&\operatorname{SINR}_{\operatorname{sum}}\\
        \operatorname{s.t.}
        & \vert\phi_{1, i}\vert=1, \quad \forall i=1,2, \ldots, N/2, \\
        & \vert\phi_{2, j}\vert=1, \quad \forall j=1,2, \ldots, N/2.
    \end{aligned}
    \label{maximizing sinr}
\end{equation}
The problem in Eq.~\eqref{maximizing sinr} is the optimization of two coupled fractional quadratic functions with unit-modulus constraints, which is non-convex and challenging to solve directly. 
To tackle this, we employ an Alternating Optimization strategy to decompose the problem into two subproblems, optimizing $\boldsymbol{\phi}_1$ and $\boldsymbol{\phi}_2$ separately. In each iteration, we fix one variable and optimize over the manifold of the other variable. The objective function is rewritten as $f(\boldsymbol{\phi}_1,\boldsymbol{\phi}_2)$, then we iteratively solve the following two subproblems until convergence: (1) For fixed $\boldsymbol{\phi}_2^{(n)}$, we solve the gradient ascent problem of $\boldsymbol{\phi}_1^{(n+1)}=\arg\max_{\boldsymbol{\phi}_1} f(\boldsymbol{\phi}_1,\boldsymbol{\phi}_2^{(n)})$ in the complex circle manifold (CCM); (2) For fixed $\boldsymbol{\phi}_1^{(n+1)}$, we solve the gradient ascent problem of $\boldsymbol{\phi}_2^{(n+1)}=\arg\max_{\boldsymbol{\phi}_2} f(\boldsymbol{\phi}_1^{(n+1)},\boldsymbol{\phi}_2)$. 

Here we take the step one as an example, the Euclidean gradient of the objective function with respect to $\boldsymbol{\phi}_1$ is
\begin{equation}
    \nabla f_1\left(\phi_1\right)=\frac{2}{C_2} \mathbf{b}_1^* \mathbf{b}_1^T \phi_1-\frac{2 C_2^{\prime}}{\left(v+\sigma^2\right)^2} \mathbf{c}_1^* \mathbf{c}_1^T \phi_1,
\end{equation}
where $C_2=\left|\mathbf{b}_2^T \phi_2\right|^2+\sigma^2 $ and $ C_2^{\prime}=\left|\mathbf{c}_2^T \phi_2\right|^2$ are constant. The gradient on the manifold is called Riemannian gradient, which can be obtained by projecting the Euclidean gradient onto the tangent space of the manifold at point. For the CCM, each point $\boldsymbol{\phi}_1$ lies on the complex unit circle, i.e., $\vert\phi_{1,i}\vert=1$. The tangent space at point $\boldsymbol{\phi}_1$ consists of all vectors $\mathbf{z}$ satisfying $\Re\{\boldsymbol{\phi}_1^* \circ \mathbf{z}\}=0$, where $\circ$ denotes the Hadamard product. The Riemannian gradient is given by
\begin{equation}
    \operatorname{grad} f_1\left(\boldsymbol{\phi}_1\right)=\nabla f_1\left(\boldsymbol{\phi}_1\right)-\Re\left\{\nabla f_1\left(\boldsymbol{\phi}_1\right) \circ \boldsymbol{\phi}_1^*\right\} \circ   \boldsymbol{\phi}_1,
\end{equation}
where $\circ$ denotes the Hadamard product. We update $\boldsymbol{\phi}_1$ along the Riemannian gradient direction with a step size $\eta$ as
\begin{equation}
    \boldsymbol{\phi}_1^{\text {temp }}=\boldsymbol{\phi}_1+\eta \operatorname{grad} f_1\left(\boldsymbol{\phi}_1\right).
\end{equation}
To ensure that the updated point remains on the CCM, we perform a retraction operation by normalizing each element of $\boldsymbol{\phi}_1^{\text {temp }}$ to unit modulus, yielding
\begin{equation}
    \boldsymbol{\phi}_1^{\text {new }}=\boldsymbol{\phi}_1^{\text {temp }}/\left|\boldsymbol{\phi}_1 ^{\text {temp }}\right|.
\end{equation}
The same procedure is applied to optimize $\boldsymbol{\phi}_2$ in the second step.
From the perspective of beamforming, the two streams can be viewed as being transmitted from two sub-MSA arrays, each responsible for directing its respective stream toward a designated receiver, as shown in Fig.~\ref{two streams}(a). By jointly optimizing phase distributions $\boldsymbol{\phi}_1$ and $\boldsymbol{\phi}_2$, each sub-array can effectively steer its beam toward its target receiver while minimizing interference to the other, as shown in Fig.~\ref{two streams}(b). This spatial separation is crucial for enhancing the overall system performance in multi-stream scenarios.

\section{MPD Metasurface Design\label{MPD Metasurface Design}}
\subsection{Unit Design: Magnitude-Phase Decoupling}
Prior research has rarely explored independent control of both magnitude and phase in metasurface unit RCs. While most existing designs operate across broadband frequencies\cite{duan2025prephase,wang2022broadband}, even narrowband implementations demonstrate different frequency selectivity that vary across phase states\cite{liao2022independent}.
To minimize out-of-band spectral pollution, the proposed design must simultaneously achieve MPD and narrowband RC magnitude ($\vert S11\vert$) selectivity.

Building upon the MSA concept and the preceding theoretical analysis of RF-domain functionalities, we establish four design guidelines for unit RC: (1)  Maximization of linear amplitude dynamic range for each phase state to enhance reflection modulation efficiency; (2) Phase response stability throughout the tunable amplitude range to maintain beam pattern consistency within each codebook configuration; (3) Ensure identical amplitude selectivity across different phase states; (4) Maintain consistent trends in amplitude variations.

\begin{figure}[ht]
    \centering
        \includegraphics[width=0.9\linewidth]{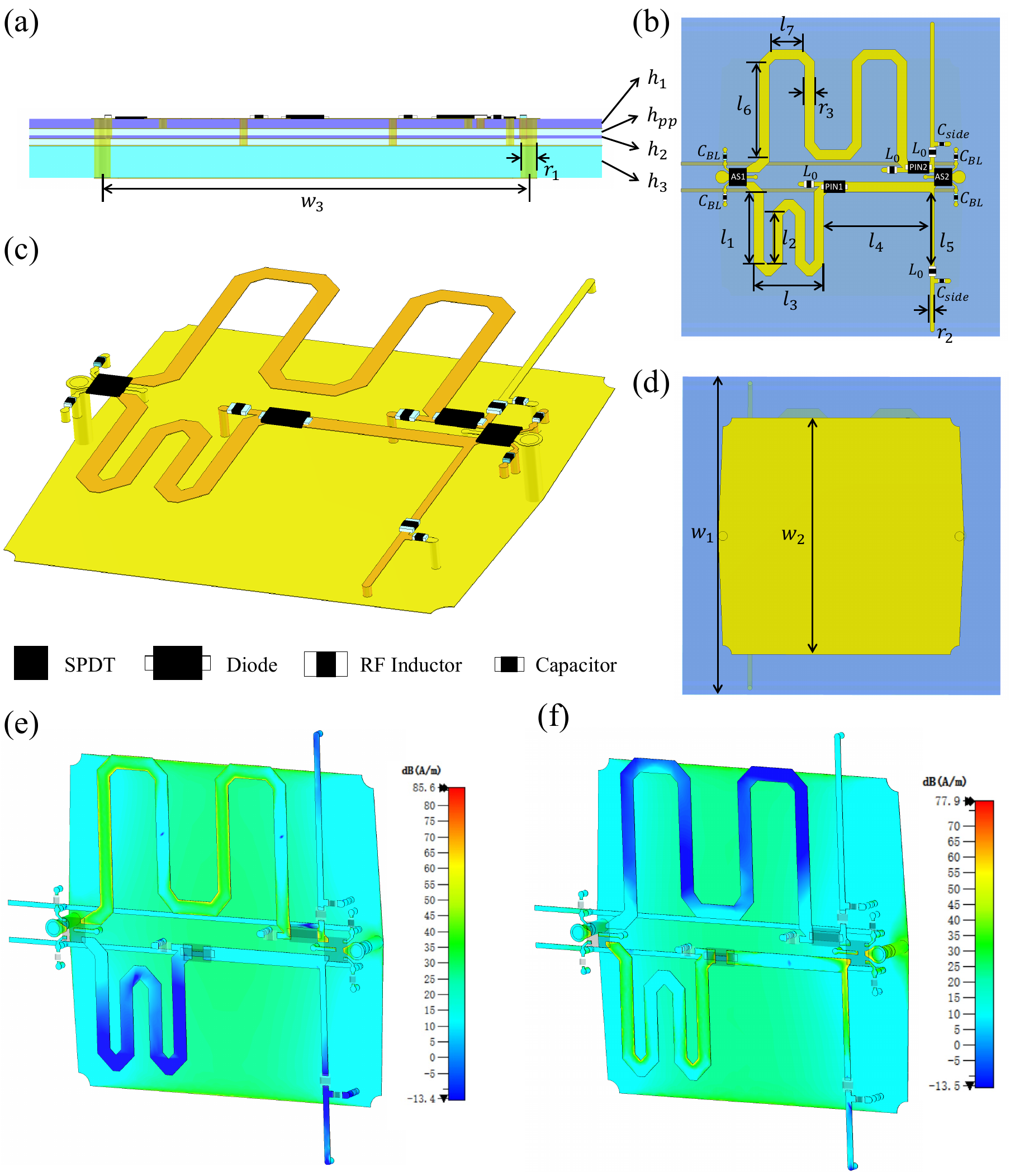}
        \caption{Hardware design of the proposed narrowband magnitude-phase decoupled unit. (a) Side view. (b) Bottom view. (c) Geometrical configuration of the metallic patch. (d) Top view. (e) Surface current distribution for State 0. (f) Surface current distribution for State 1.}
    \label{unit}
\end{figure}
Fig.~\ref{unit}(a), (b) and (d) present the side view and bottom-top view of the proposed unit cell, respectively. The geometric period is designed as $w_1=18$ mm ($w_1<\lambda/2$ at $5.8$ GHz) to suppress grating lobes in the reflected EM waves. Each unit comprises three layers, including the radiation layer, the IF signal layer, and the functional network layer. The three parts are laminated together with two
pieces of prepreg substrates. The specific dimensions of the design and the material setting can be found in Table.~\ref{Parameter}.
\begin{table}[ht]
\centering
\caption{Geometric Parameters of the Unit Cell (Unit: mm)}
\label{Parameter}
\begin{tabular}{llllllll}
\bottomrule
Para. &$l_1$  &$l_2$  &$l_3$  &$l_4$  &$l_5$  &$l_6$  &$l_7$  \\
Value     &3.818  &2.768  &1.62  &6  &4.308  &5.468  &1.8  \\ \hline
Para. &$r_1$ &$r_2$  &$r_3$  &$h_1 $ &$h_2$  &$h_3$  &$h_{pp}$  \\
Value     &0.26  &0.254  &0.56  &1  &0.1  &0.3  &0.2  \\ \hline
Para. &$w_1$  &$w_2$  &$w_3$  &$C_{\text{BL}}$  &$C_{\text{side}}$  &$L_0$  &  \\
Value     &18  &13.4  &13.4  &33  &1  &33  &  \\ \bottomrule
\end{tabular}
\end{table}
The unit adopts a dual-via coupling configuration, where the top-layer design incorporates tangential cuts along both sides of the rectangular patch adjacent to the vias. This innovative geometry significantly enhances the coupling efficiency of high-frequency energy into the underlying circuitry. The four corners of the rectangle are rounded to suppress undesired characteristic modes induced by sharp-edge radiation. 

In the functional network layer, by utilizing two SPDT RF-switches, the RF signal can be controlled to pass through either the upper path or the lower path, see Fig.~\ref{unit}(e) and (f) for the 3D simulation results of surface current.
The phase delay can be precisely controlled through careful adjustment of the electrical length in the delay path. As illustrated in Fig.~\ref{unit}(c), the upper transmission path is specifically designed with an additional $\pi$ radians (equivalent to a half-wavelength) of electrical length compared to the lower path, thereby creating the required 180° phase differential. This configuration effectively renders the SPDT RF switches functionally equivalent to a phase shifter. 

\begin{figure}[ht]
    \centering
        \includegraphics[width=0.9\linewidth]{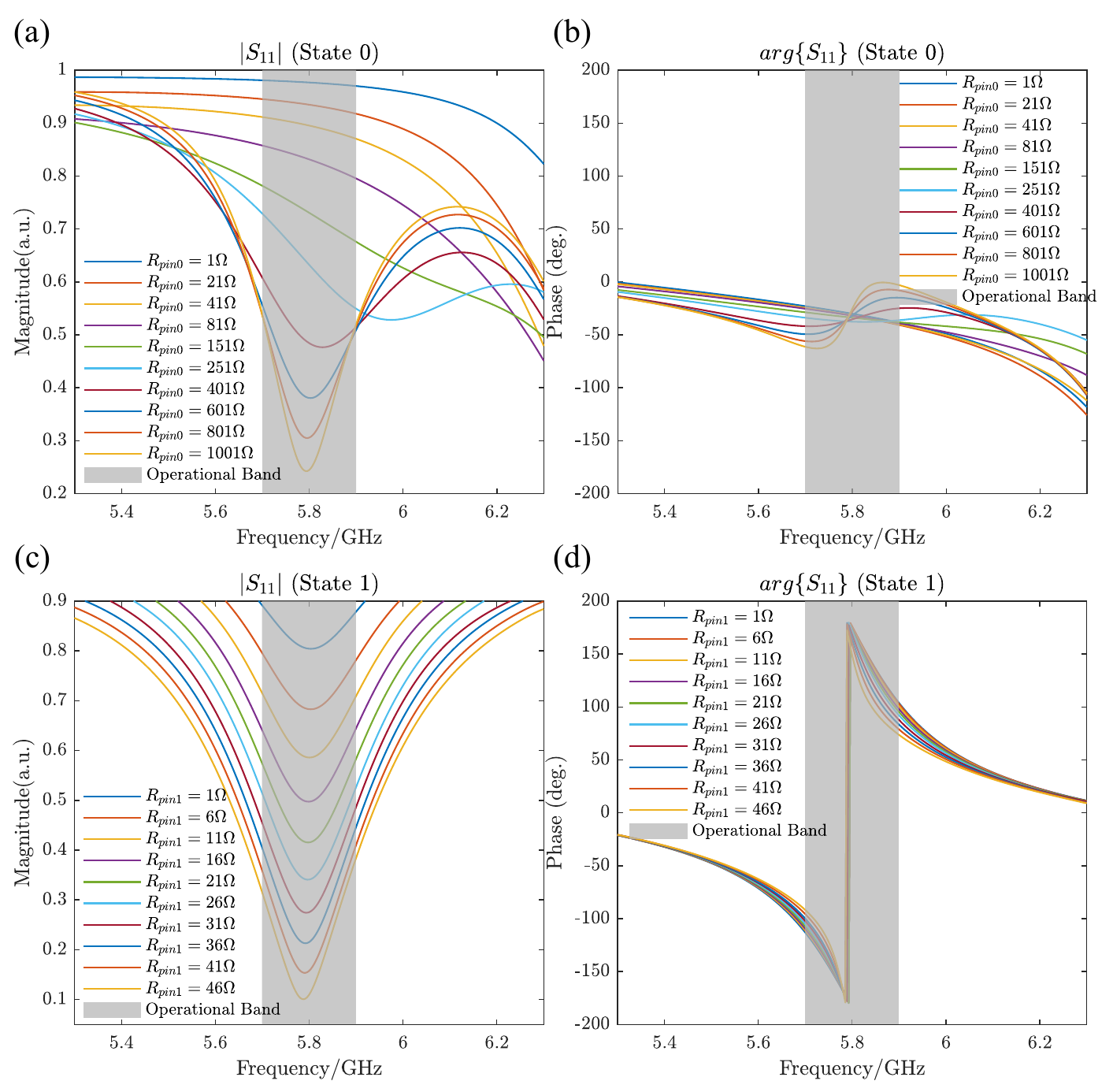}
        \caption{The simulated RC of the proposed unit cell under various bias voltages. (a) Magnitude response of State 0. (b) Phase response of State 0. (c) Magnitude response of State 1. (d) Phase response of State 1. }
    \label{S11}
\end{figure}
\begin{figure}[ht]
    \centering
        \includegraphics[width=0.8\linewidth]{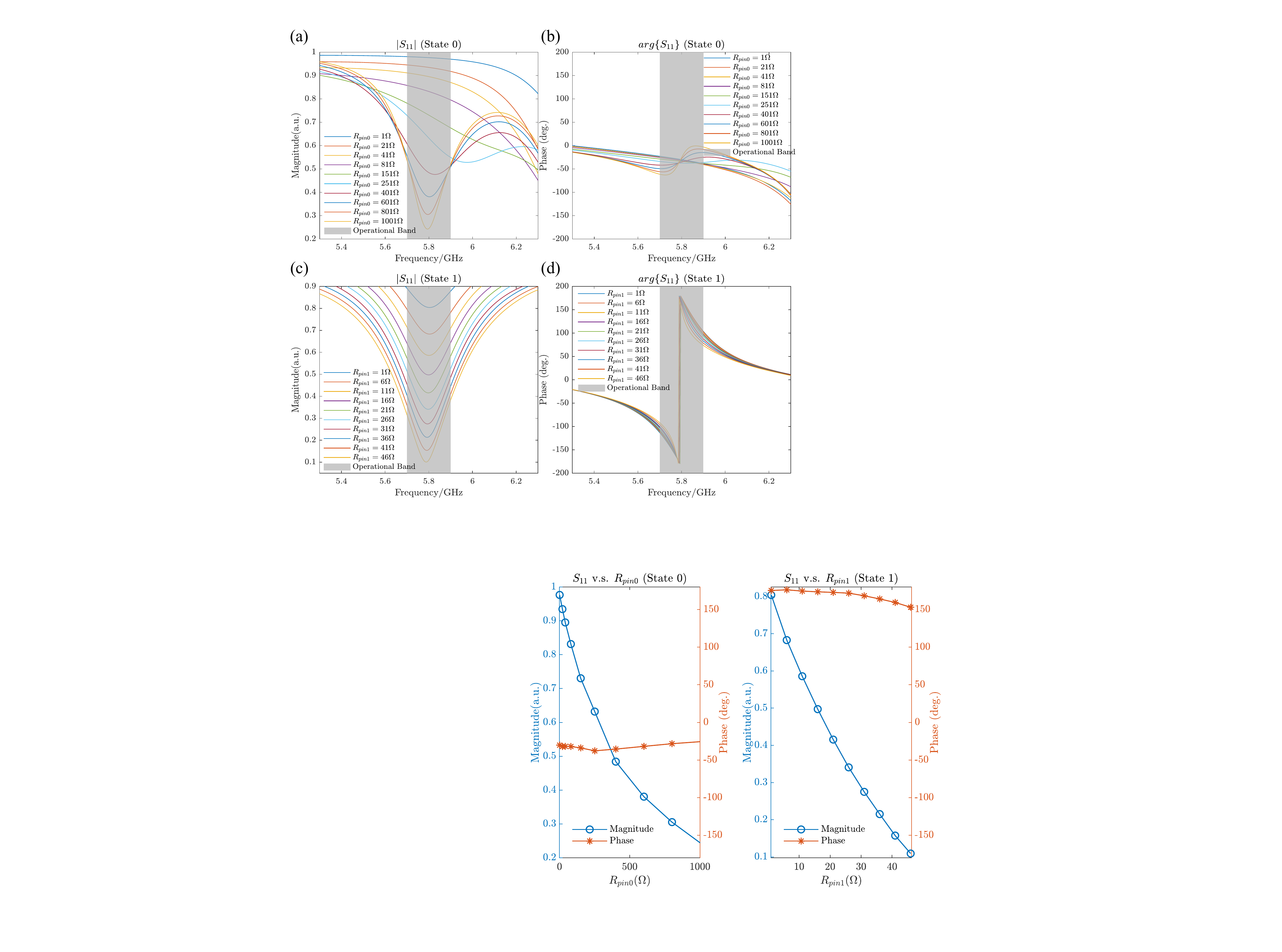}
        \caption{The measured RC of the proposed unit cell under various equivalent resistance at $5.8$ GHz.}
    \label{curve S11}
\end{figure}

As shown in Fig.~\ref{S11}, the unit cell is meticulously modeled and simulated using CST Microwave Studio 2019, with the phase and magnitude response at $5.8$ GHz. The PIN diode for state 0 (BAR64-02V) is integrated into the lower path, exhibiting an equivalent resistance range of $1$ $\Omega$ to $1001$ $\Omega$ under forward bias and a fixed parasitic inductance of $0.6$ nH. The state 1 PIN diode (SMP1345-079LF) is integrated into the upper path, with an equivalent resistance range of $1$ $\Omega$ to $46$ $\Omega$ and an identical parasitic inductance of $0.6$ nH  (see Fig.~\ref{S11}). As shown in Fig.~\ref{curve S11}, the measured results indicate the RC magnitudes ($\vert S_{11}\vert$) for the two states range from 0.8 to 0.1 and from 1 to 0.2, respectively. Furthermore, the phase responses ($\arg\{S_{11}\}$) remain stable at approximately $170^{\circ}$ and $-25^{\circ}$ for each corresponding state.

\subsection{Equivalent single-diode mixer}
\begin{figure}[ht]
    \centering
        \includegraphics[width=0.8\linewidth]{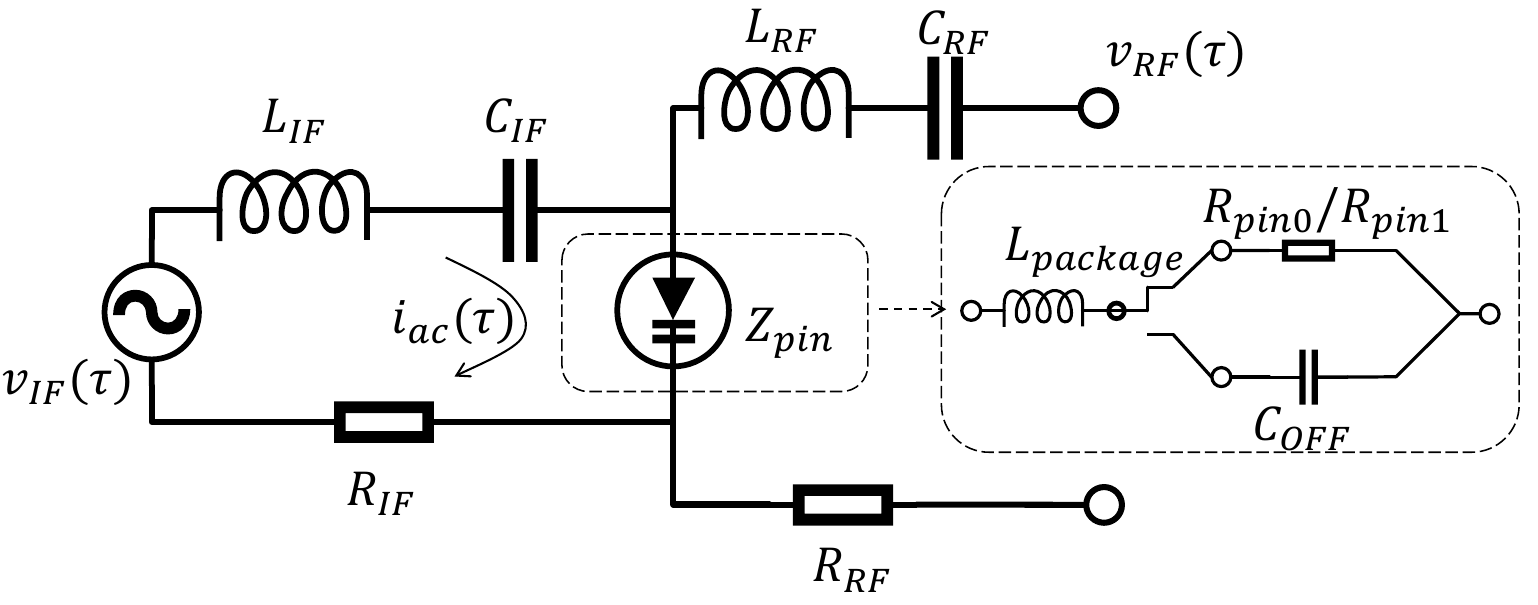}
        \caption{The equivalent circuit model of the proposed unit cell as a single-diode mixer.}
    \label{unit2}
\end{figure}

From the perspective of equivalent circuit analysis, this section elucidates the feasibility and fundamental principles of undistorted baseband signal modulation onto the incident EM wave envelope. The proposed approach employs small-signal mixers integrated into patch antenna elements utilizing PIN diodes.
Fig.~\ref{unit2} illustrates the equivalent mixer circuit representation of the proposed unit design. the variables with RF subscripts in the right-side network represent the RF equivalent circuit, while the left-side loop corresponds to the equivalent circuit for IF signals. This equivalent circuit model effectively captures the key signal conversion characteristics of the implemented mixer topology. When the diode is conducting, it can be modeled as a dynamic resistor, with its small-signal relationship expressed through a Taylor expansion as follows:
\begin{equation}
    \begin{aligned}
i(\tau) & =I_s\left(e^{\alpha\left(v_{\operatorname{RF}}(\tau)+v_{\operatorname{IF}}(\tau)\right)}-1\right) \\
& \approx I_0+\frac{v_{\operatorname{RF}}(\tau)+v_{\operatorname{IF}}(\tau)}{R_d}+\frac{\left(v_{\operatorname{RF}}(\tau)+v_{\operatorname{IF}}(\tau)\right)^2}{2 R_d^{\prime}}
\end{aligned}
\label{taylor}
\end{equation}
where $R_d$ and $R_d^{'}$ respectively denotes the dynamic equivalent resistance as shown in Fig.~\ref{unit2} (b). 
By working at the proper region of the unit, the cross-product term remains as 
\begin{equation}
    i_{a c}(\tau) \approx \frac{v_{\operatorname{RF}}(\tau) v_{\operatorname{IF}}(\tau)}{R_d^{\prime}}.
    \label{mixer}
\end{equation}

Due to the inherent nonlinearity in the diode's junction resistance, equivalently the nonlinearity of the RC, two superimposed input signals driving the unit can result in the multiplication of these two signals. The conclusion is that by utilizing the dynamic junction resistance characteristics of the forward-biased diodes on the metasurface, IF signals can be modulated onto the outgoing EM waves. 
Moreover, to avoid signal distortion, we need to select a working range where $R_d'$ remains close to a constant, which means the first-order derivative of junction resistance $R_d^{'}$ remains relatively stable, ensuring undistorted modulation signals (see Fig.~\ref{nonlinear distortion}). 
This criterion provides essential guidance for determining a suitable amplitude range for $v_{\operatorname{IF}}(\tau)$ and for performing calibration to compensate for residual nonlinearity.
\section{Experimental Validation\label{Experimental Validation}}
We evaluated the multifunctional capabilities of the MSA architecture through a series of experiments in two representative application scenarios: metasurface-based wireless communications and radar sensing. 
\subsection{MSA Prototype}
\begin{figure}[ht]
    \centering
        \includegraphics[width=\linewidth]{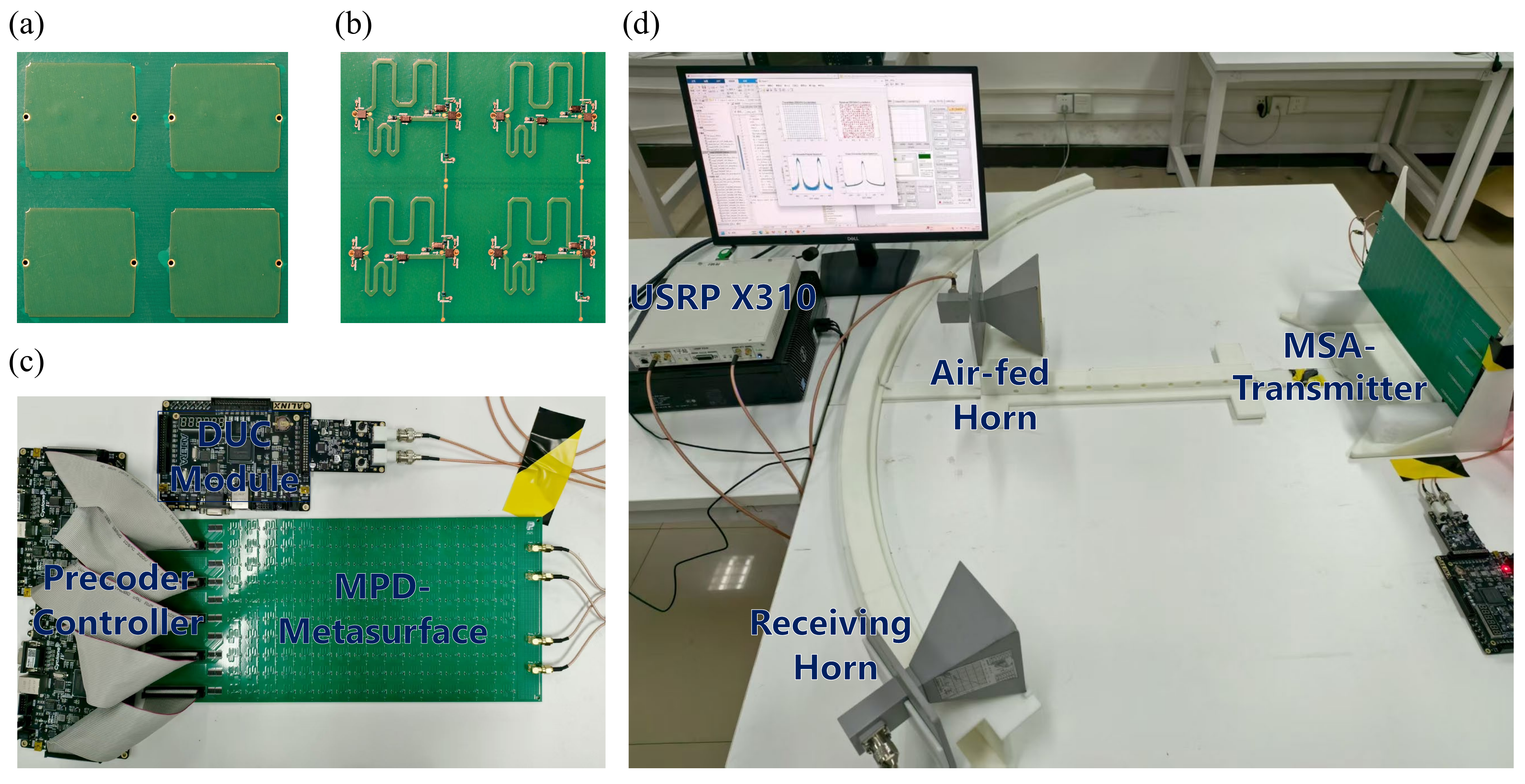}
        \caption{(a) The picture of the top view of unit cells. (b) The picture of the bottom view of unit cells. (c) The entire MSA prototype system.}
    \label{prototype}
\end{figure}

The proposed MSA transmitter operates at a center frequency of $f_c =5.8$ GHz and comprises a $16\times 10$ array of MPD unit cells. The metasurface was designed using CST Microwave Studio and fabricated using standard printed circuit board technology. All experimental data presented in the preceding sections were collected using a custom-built test platform incorporating a software-defined radio device (i.e., NI USRP-X310) and a post-processing computer.
The prototype (see Fig.\ref{prototype}(c)) consists of the MPD-Metasurface, the beamforming controller (two AX515 FPGAs) and  the DUC module (one AN9767 DAC and one AX515 FPGA). For the sake of convenience, we used several splitters for ensuring the input voltage signal is identical. The integration of two diode types within each unit cell results in partially overlapping linear operating regions. To maximize system efficiency, we implemented dedicated IF control interfaces for each diode type.
\subsection{Symbol-level Isotropic MSA-Backscatter System}
\begin{figure*}[ht]
    \centering
        \includegraphics[width=\linewidth]{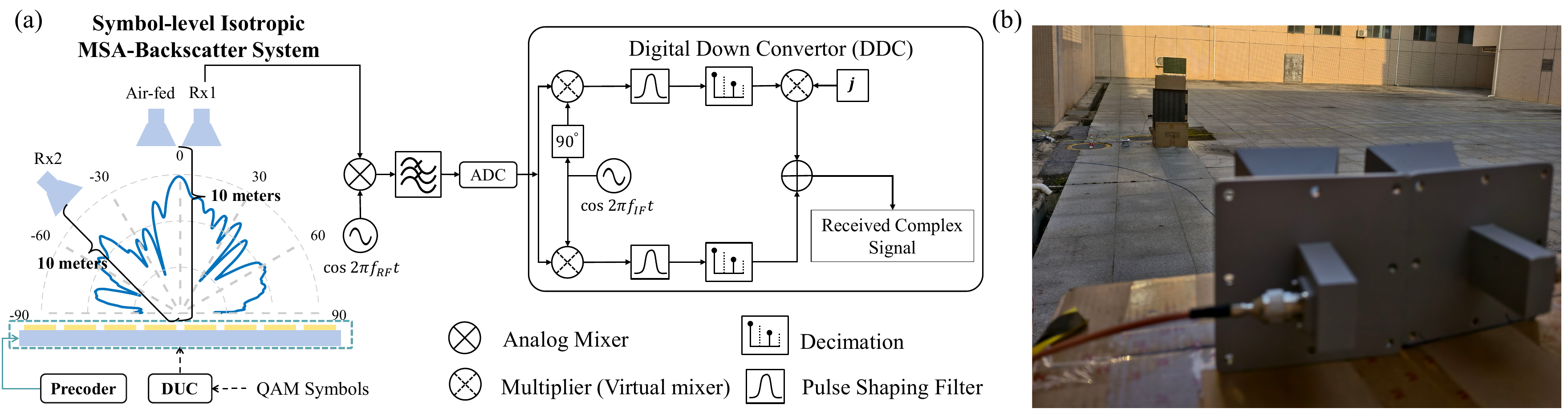}
        \caption{(a) The schematic diagram of the entire system including the procedures of DUC and DDC. (b) Experimental scenario. }
    \label{MSA-Backscatter-schematic}
\end{figure*}

To validate the proposed MSA-transmitter, a realistic wireless communication system was built to perform experiments of backscatter data transmission in outdoor environment as shown in Fig.~\ref{MSA-Backscatter-schematic}(b). 
The schematic of the entire system is illustrated in Fig.~\ref{MSA-Backscatter-schematic}(a). The MSA transmitter is illuminated by a $5.8$ GHz continuous wave carrier signal generated by the USRP through a transmitting horn antenna. The reflected signal from the MSA is received by two positionally-distinct horn antennas connected to another USRP for signal demodulation and performance evaluation. The distance between the MSA and the transmitting antenna is approximately $10$ m, while the distance to each receiving antenna is around $10$ m. The two receiving antennas are separated by an angle of about $45^{\circ}$ relative to the MSA center.
\begin{figure}[h]
    \centering
        \includegraphics[width=0.8\linewidth]{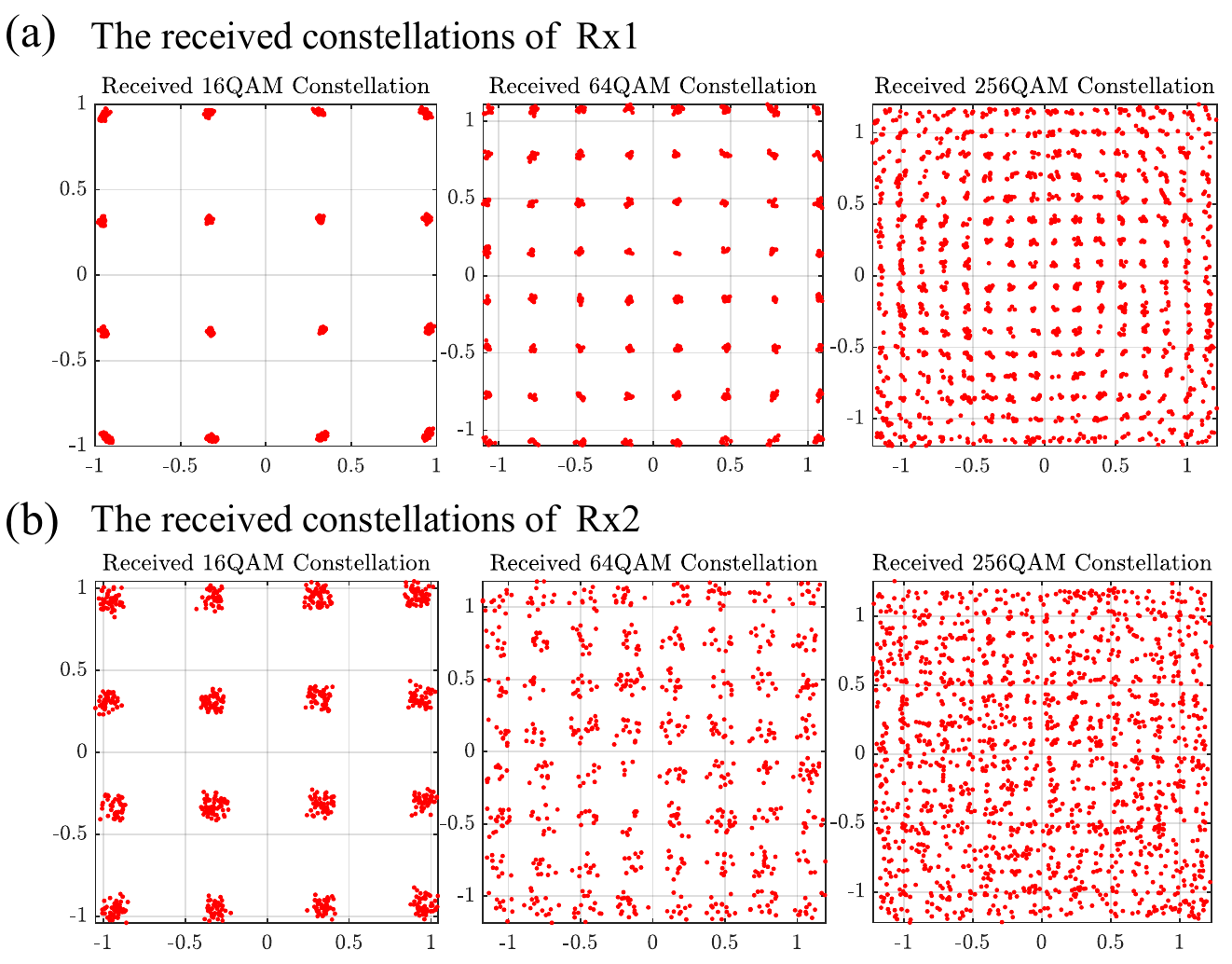}
        \caption{(a) The received constellation diagrams at Rx1 for different modulation orders, ranging from 16-QAM to 256-QAM. (b) The received constellation diagrams at Rx2 for different modulation orders, ranging from 16-QAM to 256-QAM.}
    \label{MSA-Backscatter}
\end{figure}
At the receiver side, the incoming signal first undergoes down-conversion through an RF mixer. It is then sampled by an ADC at the same sampling rate to obtain the received digital IF signal. Subsequently, through digital down-conversion (DDC) (the inverse process of DUC), the original baseband complex symbols can be reconstructed. Figs. \ref{MSA-Backscatter}(a) and (b) present the received constellations at Rx1 and Rx2, respectively, for various QAM modulation orders ranging from 16-QAM to 256-QAM. This experiment demonstrates that the MSA wireless communication system is capable of generating and transmitting signals of arbitrary high orders while maintaining spatial symbol-level isotropy.

\subsection{Diversity and Interference Cancellation Validation}
\begin{figure}[h]
    \centering
        \includegraphics[width=0.7\linewidth]{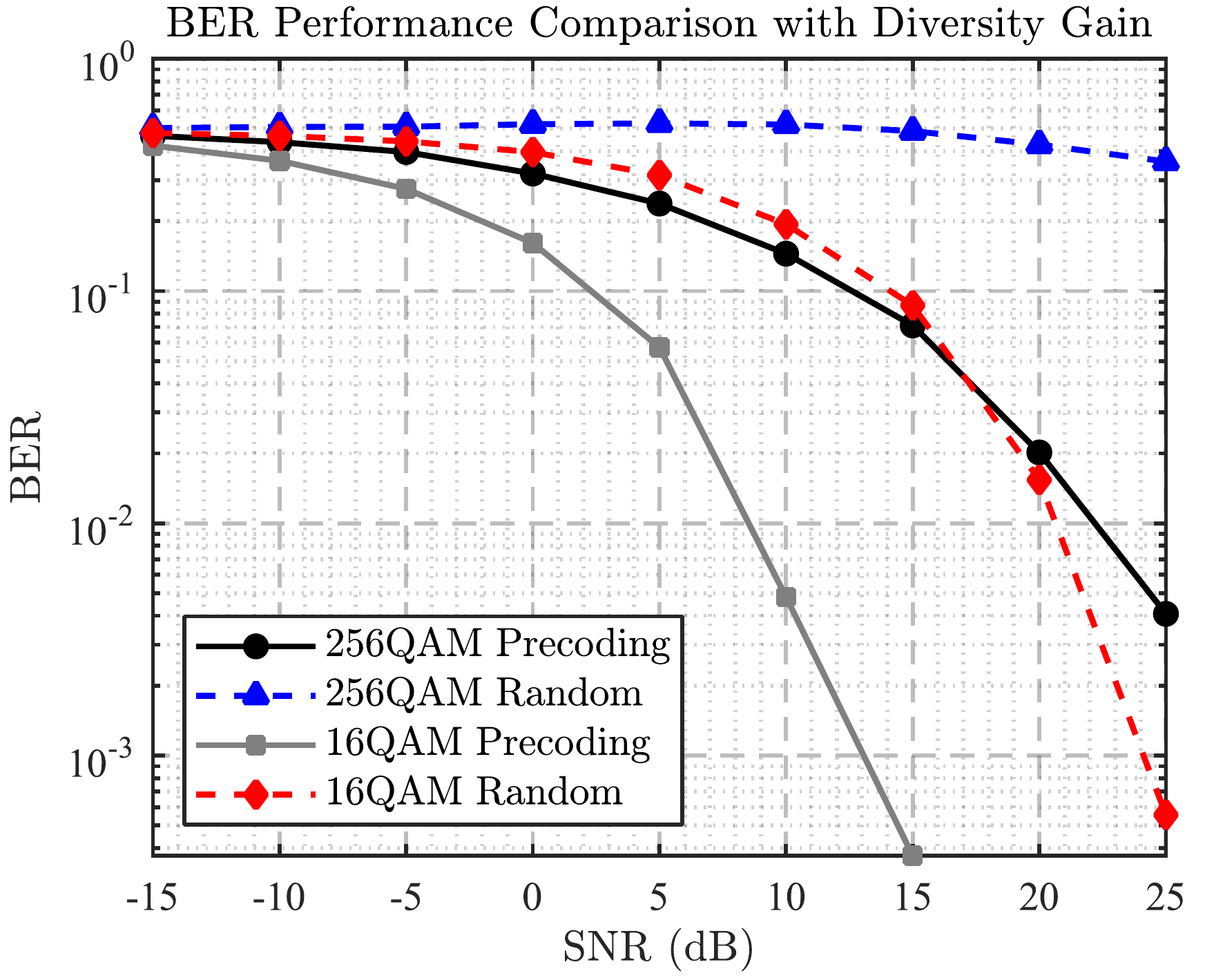}
        \caption{The experimental BER performance of MSA-transmitter with and without diversity gain under 16-QAM and 256-QAM scheme.}
    \label{diversity}
\end{figure}

\begin{figure}[h]
    \centering
        \includegraphics[width=0.8\linewidth]{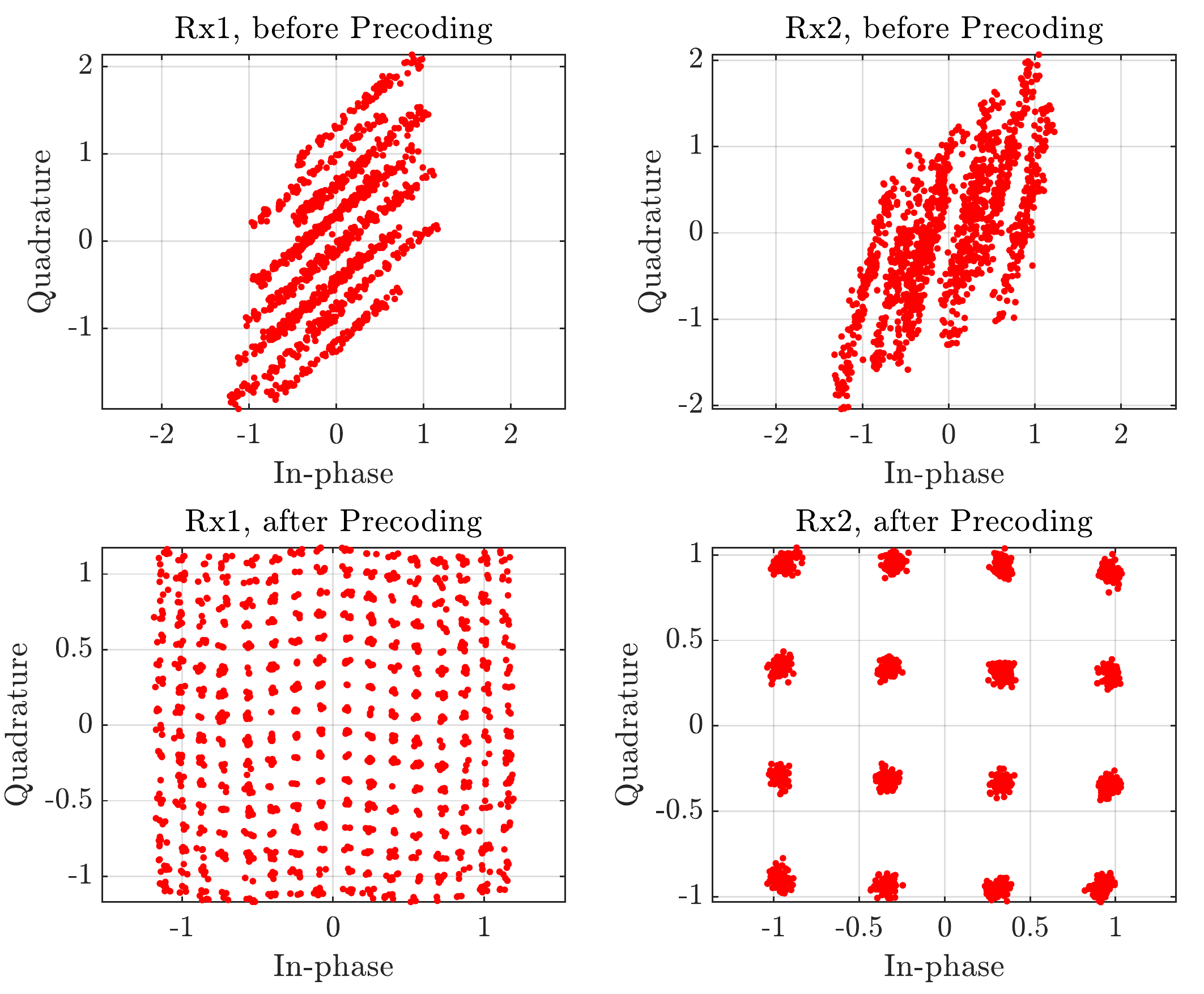}
        \caption{The experimental constellation diagrams for two receivers (Rx1 and Rx2) before and after interference cancellation.}
    \label{two streams results}
\end{figure}

This subsection presents experimental validation of the MSA architecture's capabilities in achieving diversity gain and effective interference cancellation in multi-stream transmission scenarios. The results, summarized in Figs~\ref{diversity} and Figs~\ref{two streams results}, demonstrate the system's performance under practical operating conditions. Figs~\ref{diversity} shows the BER performance comparison for 256-QAM and 16-QAM modulation schemes with and without precoding. The results clearly demonstrate that the proposed precoding scheme achieves significant diversity gain across both modulation orders. For 16-QAM, the diversity gain is maintained at about 10dB SNR improvement. This consistent performance improvement validates the effectiveness of the proposed optimal phase optimization (Section~\ref{diversity section}) for one-stream transmission in enhancing link reliability through spatial diversity.

The interference cancellation capability of the MSA architecture is evaluated in Fig.~\ref{two streams results}, which presents constellation diagrams for two separated receivers (Rx1 and Rx2) before and after applying the precoding scheme proposed in Section~\ref{multi streams}. For the convenience of demonstration, the modulation orders of two receivers are set as different. Before precoding, both receivers experience severe inter-stream interference, as evidenced by the significantly distorted constellations. After applying the optimized precoding matrices, both streams exhibit well-defined constellation points with minimal inter-symbol interference. The clear separation of constellation clusters in the "After Precoding" subplots demonstrates the MSA's ability to effectively cancel inter-stream interference while maintaining signal integrity. 
These experimental results collectively validate that the MSA architecture successfully addresses two critical challenges in multi-stream metasurface transmitters: achieving reliable diversity gain and effective interference cancellation. The demonstrated performance highlights the practical viability of the proposed architecture for next-generation wireless communication systems requiring high spectral efficiency and multi-user support.

\subsection{MSA-Doppler-Spoofing System}
The MSA offers significant advantages as a radar deception tag, including the ability to generate arbitrary high-precision spectra and time-frequency Doppler signatures, minimizing interception risk by eliminating leakage from sidelobe variation patterns. Additionally, it enables simultaneous generation of specific time-frequency signatures while performing beamforming to enhance or reduce the radar cross-section as required.
Metamaterials or metasurface devices are often used as radar stealth coatings or deception tags \cite{chen2014reduction,kozlov2023radar}. With the advent of information metasurfaces, the flexibility and reconfigurability of metasurfaces as radar deception tags have significantly improved. Existing information metasurfaces, when used as Doppler radar deception reflection tags, face two difficult challenges: (1) Due to signal-beampatterns coupling, the spectrogram differ at various positions, making it easy to detect that the forged signal is artificial; (2) The unknown position of the radar makes it difficult to present the radar with the desired custom waveform. These two major issues greatly limit the practical use of metasurfaces as Doppler-spoofing tags.
The introduction of MSA can address these two problems. 

\begin{figure}[ht]
    \centering
        \includegraphics[width=\linewidth]{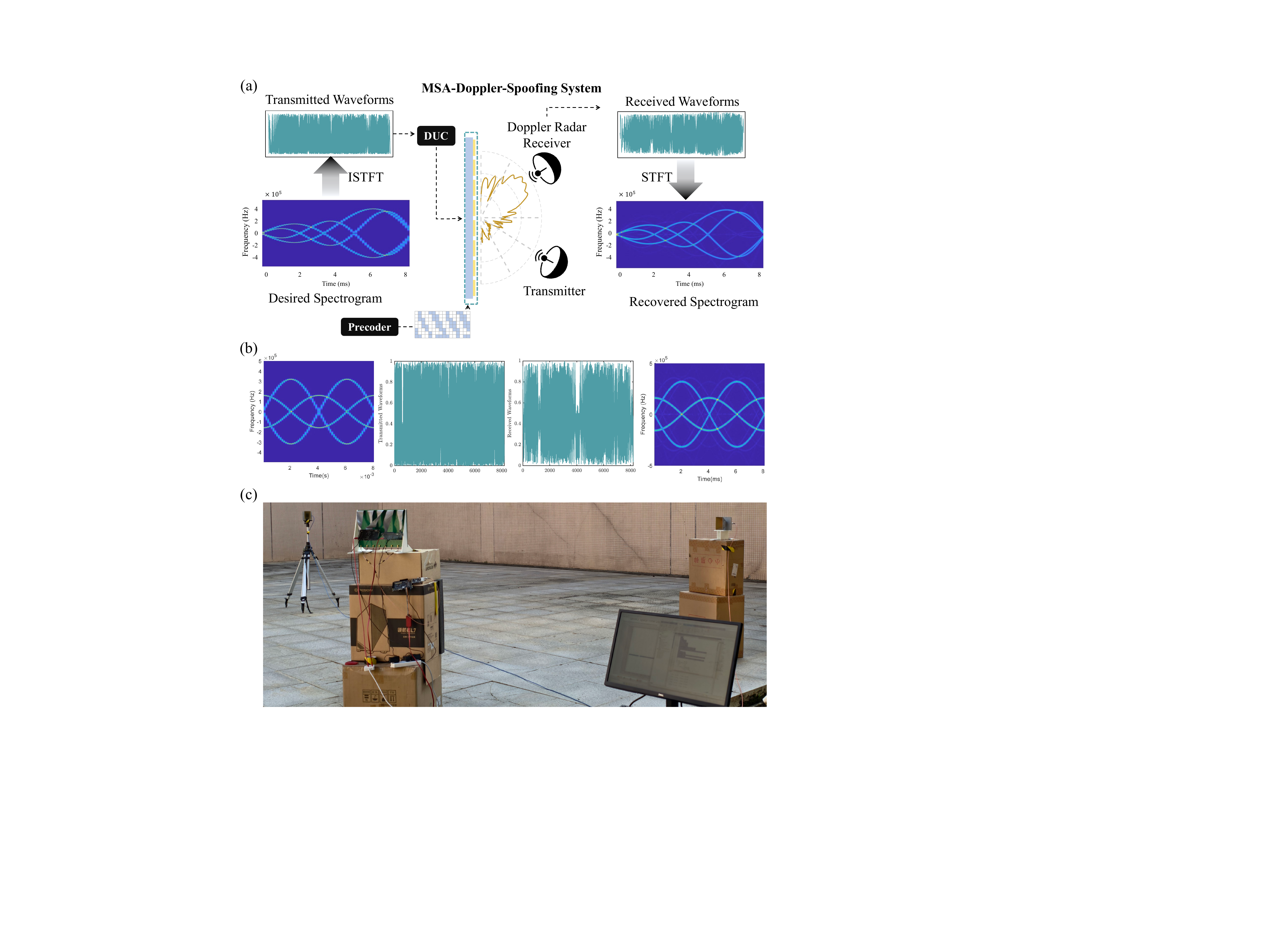}
        \caption{(a) Schematic diagram of MSA-Backscatter system and experimental received spectrogram. (b) The four experimental data diagrams respectively denote the desired spectrogram of a dual rotor helicopter, transmitted waveform, received waveform and recovered spectrogram}
    \label{MSA-Doppler-Spoofing}
\end{figure}

As shown in Fig.\ref{MSA-Doppler-Spoofing}, by performing an inverse short-time Fourier transform on the desired spectrogram, the time-domain waveform can be obtained, and the corresponding waveforms can be input into the MPD-Metasurface through a DAC. The MSA can generate the desired time-frequency signature at different observation angles, effectively eliminating the risk of being detected as an artificial signal. Furthermore, by leveraging the spatial degrees of freedom provided by the MSA architecture, beamforming can be performed to direct the forged signal towards the radar's location, enhancing the effectiveness of the deception.

\subsection{Nonlinear distortion analysis}
\begin{figure}[ht]
    \centering
        \includegraphics[width=0.8\linewidth]{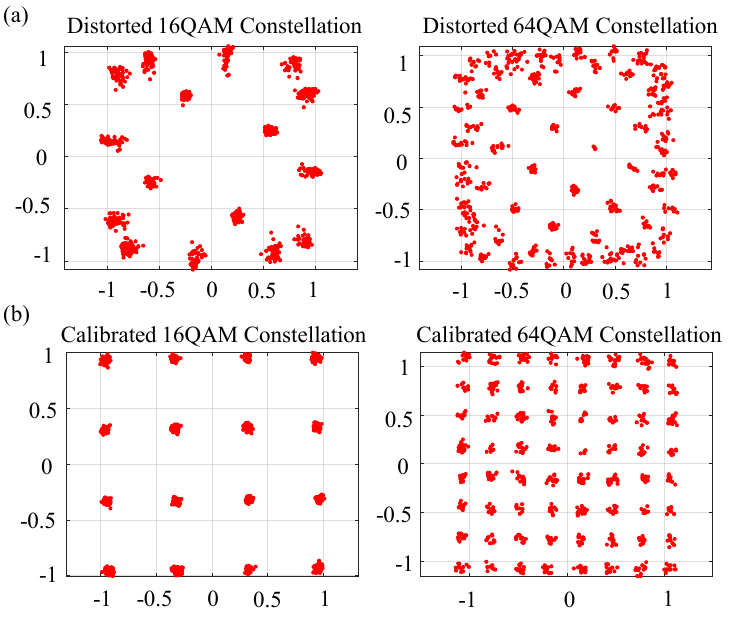}
        \caption{(a) Distorted constellations diagrams of 16-QAM and 64-QAM. (b)  Calibrated constellations diagrams of 16-QAM and 64-QAM.}
    \label{nonlinear distortion}
\end{figure}

As discussed in the previous section, to avoid signal distortion during the modulation process, it is crucial to select an appropriate operating range where the first-order derivative of the junction resistance \(R_d'\) remains relatively stable. This ensures that the modulation signals are transmitted without distortion.
To validate this principle, we conducted experiments to analyze the nonlinear distortion characteristics of the MSA transmitter. As shown in Fig.~\ref{nonlinear distortion}(a), when the diode voltage progressively exceeds its linear operating range, significant nonlinear distortion occurs in the output signal. This distortion manifests as a deviation from the intended waveform, leading to performance degradation in the error vector magnitude (EVM) and BER.
Conversely, as illustrated in Fig.~\ref{nonlinear distortion}(b), when the diode voltage is maintained within the specified linear operating range, the output signal closely matches the desired waveform, exhibiting minimal distortion. This results in improved EVM and BER performance, confirming the effectiveness of operating within the linear region of the diode's junction resistance.

\subsection{Data rate limit testing of Prototype}
To evaluate the data rate limit of the MSA prototype, we conducted a series of experiments varying the sample rate, IF, and samples per symbol. The results are summarized in Table~\ref{rate testing}. The experiments demonstrate that as the sample rate and symbol rate increase, the achievable data rate also increases. However, this comes at the cost of a higher BER, indicating a trade-off between data rate and transmission reliability. The highest data rate achieved in our tests was 20 Mbps with a BER of $1.98\times10^{-2}$ at a sample rate of 20 MHz and an IF of 5 MHz, without error correction coding.

\begin{table}[ht]
    \caption{Data rate testing for 256-QAM transmission without error correction coding}
    \centering
    \label{rate testing}
\begin{tabular}{cccccc}
\bottomrule
\begin{tabular}[c]{@{}c@{}}Sample \\ Rate\\ (MHz)\end{tabular} & \begin{tabular}[c]{@{}c@{}}IF \\ (MHz)\end{tabular} & \begin{tabular}[c]{@{}c@{}}Samples\\ per\\ Symbol\end{tabular} & \begin{tabular}[c]{@{}c@{}}Symbol\\  Rate\\ (M/s)\end{tabular} & \begin{tabular}[c]{@{}c@{}}Data\\ Rate \\ (Mbps)\end{tabular} & BER    \\ \hline
2& 0.5& 10& 0.2& 1.6& $1\times10^{-4}$ \\
10& 2.5& 10& 1& 8& $7\times10^{-4}$ \\
10& 2.5& 8& 1.25& 10& $3.9\times10^{-3}$ \\
20& 5& 8& 2.5& 20& $1.98\times10^{-2}$ \\ \bottomrule
\end{tabular}
\end{table}

\section{Conclusion\label{Conclusion}}
This paper has presented a novel MSA that fundamentally transforms the design of metasurface transmitters. By introducing a hardware-decoupled, dual-stage up-conversion process, the MSA successfully overcomes the fundamental limitations of conventional designs, namely modulation-order restriction, symbol anisotropy, and harmonic interference.
Experimental results from a 5.8 GHz prototype validate the architecture's superiority. The system demonstrates spatially isotropic 256-QAM transmission and high-fidelity waveform generation for joint communication and sensing, achieving data rates up to 20 Mbps.
The MSA establishes a practical and scalable foundation for next-generation wireless systems. Future work will focus on extending the architecture to broadband operation and integrating real-time adaptive precoding for dynamic environments, further advancing its application in 6G multi-functional networks.

\appendices



\ifCLASSOPTIONcaptionsoff
  \newpage
\fi

\bibliographystyle{IEEEtran}

\end{document}